\documentclass[12pt,prb]{revtex4}
\makeatletter
\def\@dotsep{4.5}
\makeatother

\usepackage{amsmath}
\usepackage{graphicx}

\begin{document}

\begin{titlepage}
\title{Polymer translocation through pores with complex geometries}
\author{Aruna Mohan}
\author{Anatoly B. Kolomeisky}
\author{Matteo Pasquali}
\affiliation{Department of Chemistry and Department of Chemical and Biomolecular Engineering, Rice University,
Houston, TX 77005}
\date{\today}

\begin{abstract}
We propose a method for the theoretical investigation of polymer translocation through composite pore structures possessing arbitrarily specified geometries. Translocation through each constituent part of the composite is treated as being analogous to the diffusion of the translocation coordinate over the free energy landscape derived from the chain configurations within the pore. The proposed method accounts for possible reverse motions of the leading chain end at the interface between constituent parts of a composite pore, a possibility that has been neglected in prior studies. As an illustration of our method, we study the translocation of a Gaussian chain between two spherical compartments connected by a cylindrical pore, and by a composite pore consisting of two connected cylinders of different diameters, which is structurally similar to the $\alpha$-hemolysin membrane channel. We demonstrate that reverse chain motions between the pore constituents may contribute significantly to the total translocation time. Our results further establish that translocation through a two-cylinder composite pore is faster when the chain is introduced into the pore on the \emph{cis} (wide) side of the channel rather than the \emph{trans} (narrow) side.
\end{abstract}

\maketitle
\end{titlepage}

\section{\label{sec:intro} Introduction}

The migration of biopolymers through nanopores plays a key role in several biological processes, including the transport of mRNA molecules from the nucleus to the cytoplasm following transcription, the transport of proteins to and from the nucleus, and the injection of viral DNA into a host cell \cite{lodish}. Biopolymer translocation further has technological applications in the development of biosensors for polynucleotide analysis and sequencing \cite{kasiano, akeson, mellerah, vercout, dekker}. In view of its biological and technological significance, polymer translocation has been the subject of several recent experimental \cite{kasiano, akeson, mellerah, henrickson, mellerprl, vercout, mathe, storm, dekker}, theoretical \cite{sung, park, muthu, lubensky, kardar1, ambjornsson, muthujcp03, slonkina, metzler, kardar2, kotsev, wm} and simulation-based \cite{muthuMC, kong2, makarov1, makarov2, matysiak, kong, forrey} investigations. These, as well as related studies concerning polymer translocation phenomena, are summarized in the review of Meller \cite{meller}.

While many prior theoretical investigations have focused on translocation through a narrow pore embedded in a rigid wall of negligible thickness, wherein the pore is assumed to accommodate at most a single polymer segment \cite{sung, muthu}, pores in biological systems can be significantly large. Recently, there has been much interest in examining the effect of pore geometry on the translocation process. Studies in this direction include investigations into the influence of pore length on the translocation time \cite{slonkina} and the release of a chain from one spherical vesicle into another \cite{park}. Pore structure is expected to play a significant role in controlling translocation through membrane channels in biological systems. In particular, experimental studies have shown that translocation occurs more readily when the chain is introduced through the side of the asymmetric $\alpha$-hemolysin membrane channel that has the larger vestibule \cite{henrickson}. An investigation of the influence of pore geometry on translocation dynamics entails the detailed consideration of chain configurations within the pore, as well as chain motions through the constituents of a composite pore structure. To this end, Muthukumar \cite{muthujcp03} has considered polymer translocation through several pore geometries, including the migration of a chain from one sphere to another through a narrow cylindrical pore. The latter study was further generalized by Wong and Muthukumar \cite{wm}, who took account of the finite diameter of the cylindrical pore by allowing the number of segments within the pore to vary and explicitly determining the corresponding free energies of the confined chain.

The works of Muthukumar \cite{muthujcp03} and Wong and Muthukumar \cite{wm} provide a framework for the explicit treatment of polymer translocation through composite pore structures with account for all possible chain configurations within the pore. However, the assumptions underlying these studies do not allow for the unrestricted motion of the chain at the interface between adjacent constituent parts of the pore. Specifically, during translocation from the donor sphere to the receptor sphere through a cylindrical pore, the leading chain end was assumed to be ``absorbed'' at the cylinder--receptor sphere interface immediately upon its arrival at the cylinder end, and reverse chain motions at the interface were disallowed \cite{muthujcp03}. A subsequent attempt was made to relax this assumption, by allowing a small, fixed number of chain segments at the leading end to retreat into the cylinder even after entering the receptor sphere \cite{wm}. Nonetheless, possible chain motions at the interface continue to be curtailed under this assumption, whereas, in reality, the occurrence of reverse chain motions may contribute significantly to the total translocation time.

In the present contribution, we propose a modification of the method of Muthukumar and coworkers \cite{muthujcp03, wm} whereby the leading chain end is allowed to move backward as well as forward at the interface between adjacent parts of the pore, and may arrive at the interface any number of times before chain migration into the pore constituent immediately downstream is initiated. This is achieved by imposing the more realistic radiation boundary condition \cite{szabo} in place of the absorbing boundary condition on the translocation coordinate at the interfaces between constituent parts of the pore. In the case of chain transport from one sphere to another through a cylindrical pore, we demonstrate that backward chain motions may lead to a significant increase in total translocation time. Our results indicate that the translocation time increases monotonically with increase in cylinder length, rather than exhibiting a minimum with respect to the cylinder length as predicted by Wong and Muthukumar \cite{wm}. We next consider the translocation of a chain from one sphere to another through a composite two-cylinder pore, which serves as a simple model of the  mushroom-shaped $\alpha$-hemolysin membrane channel \cite{ah}. Our results confirm that translocation is faster when the chain is introduced into the pore on the \emph{cis} (wide) side of the channel rather than the \emph{trans} (narrow) side.

The paper is organized as follows. In Sec. \ref{sec:cylinder}, we treat the translocation of a Gaussian chain from the donor sphere to the receptor sphere through a cylindrical pore. Translocation through an $\alpha$-hemolysin pore is considered in Sec. \ref{sec:ah}, while Sec. \ref{sec:sum} provides a summary of our findings. 

\section{\label{sec:cylinder} Two spheres connected by a cylinder}

\subsection{Free energy landscape}

We first briefly summarize the approach previously employed by Park and Sung \cite{park} and Muthukumar and coworkers \cite{muthujcp03,wm} and adopted here by us for the derivation of the free energy landscape for the translocation process. The probability density $P(\mathbf{r}, \mathbf{r}_0, N)$ that the ends of a Gaussian chain of $N$ segments are at positions $\mathbf{r}$ and $\mathbf{r}_0$ is governed by the equation \cite{doi}
\begin{equation}
\left( \frac{\partial }{\partial N}-\frac{1}{6}\nabla _{\mathbf{r}%
}^{2}\right) P(\mathbf{r},\mathbf{r}_{0},N)=0 \label{eq:govern}
\end{equation}
where the unit of length has been set equal to the Kuhn length (denoted $b$ by Wong and Muthukumar). Equation (\ref{eq:govern}) is solved subject to the condition $P(\mathbf{r},\mathbf{r}_{0},0)=\delta ^{3}\left( \mathbf{r}-\mathbf{r}_{0}\right)$ and the condition that $P(\mathbf{r},\mathbf{r}_{0},N)$ vanish at all surfaces.

The solution of Eq. (\ref{eq:govern}) in conjunction with the appropriate boundary conditions for a chain confined to a sphere of radius $R$  yields the expression
\begin{equation}
P_{S}(r,r_{0},N)=\frac{1}{2\pi R r r_{0}}\sum_{m=1}^{\infty }\sin \left( \frac{%
m\pi r}{R}\right) \sin \left( \frac{m\pi r_{0}}{R}\right) \exp \left(
-\frac{m^{2}\pi ^{2}}{6R^{2}}N\right) \label{eq:PS}
\end{equation}
under the assumption that the probability density depends only on the magnitudes $r$ and $r_0$ \cite{muthujcp03,wm}, with the origin chosen to lie at the sphere center. If one chain end, say, $\mathbf{r}_0$, is allowed to lie anywhere within the sphere while the other end, $\mathbf{r}$, is tethered arbitrarily close to the sphere surface such that $r=R-c$, with $c$ the tethering distance, then the probability density per unit area of the sphere surface corresponding to $\mathbf{r}$ becomes
\begin{equation}
P_{S_{1}}(R,N)=\frac{2c}{R}\sum_{m=1}^{\infty }\exp \left(%
\frac{ -m^{2}\pi ^{2}}{6R^{2}}N\right) \label{eq:PS1}
\end{equation}

Similarly, we obtain the expression
\begin{multline}
P_C(\mathbf{r},\mathbf{r}_{0},N)=\frac{4}{\pi Ma^{2}}\sum_{m=1}^{\infty
}\sum_{k=1}^{\infty }\sum_{n=0}^{\infty }\left[ \frac{1}{2}\delta
_{n0}+\left( 1-\delta _{n0}\right) \left( \cos (n\theta )\cos (n\theta
_{0})+\sin (n\theta )\sin (n\theta _{0})\right) \right] \\
\times \frac{J_{n}\left( \frac{\mu _{n,k}r}{a}\right) J_{n}\left( \frac{\mu
_{n,k}r_{0}}{a}\right) }{J_{n+1}^{2}\left( \mu _{n,k}\right) }\sin \left[ 
\frac{m\pi }{M}\left( z+\frac{M}{2}\right) \right] \sin \left[ \frac{m\pi }{M%
}\left( z_{0}+\frac{M}{2}\right) \right] \exp \left[ -\left( \frac{\mu
_{n,k}^{2}}{a^{2}}+\frac{m^{2}\pi ^{2}}{M^{2}}\right) \frac{N}{6}\right] \label{eq:PC}
\end{multline}
for a chain confined to a cylinder of length $M$ and radius $a$, and whose ends lie at positions $\mathbf{r}=(r,\theta,z)$ and $\mathbf{r}_0=(r_0, \theta_0, z_0)$ measured relative to an origin located at the center of the cylinder axis. If one chain end, $\mathbf{r}_0$, is free and the end at $\mathbf{r}$ is fixed such that $z=-M/2+c$, where the $z$-coordinate is measured along the cylinder axis and $c$ is a small tethering distance, then
\begin{equation}
P_{C_{1}}(N)=\frac{8\pi a^{2}c}{M}\sum_{m,k=1}^{\infty }\frac{1-(-1)^{m}}{\mu
_{0,k}^{2}}\exp \left[ -\left( \frac{\mu _{0,k}^{2}}{a^{2}}+\frac{m^{2}\pi
^{2}}{M^{2}}\right) \frac{N}{6}\right] \label{eq:PC1}
\end{equation}
For a chain whose ends are fixed such that $z=-M/2+c_1$ and $z_0=M/2-c_2$, we obtain the expression
\begin{equation}
P_{C_{2}}(N) = \frac{8\pi ^{3}a^{2}c_{1}c_{2}}{M^{3}}\sum_{m,k=1}^{\infty }%
\frac{m^{2}}{\mu _{0,k}^{2}}\left( -1\right) ^{m+1}\exp \left[ -\left( \frac{%
\mu _{0,k}^{2}}{a^{2}}+\frac{m^{2}\pi ^{2}}{M^{2}}\right) \frac{N}{6}\right] \label{eq:PC2}
\end{equation}
The above expressions are correct to leading order in the tethering distances, and the symbols $\mu_{0,k}$, $k=1,2,...$ appearing therein are the roots of the Bessel function of the first kind of order $0$, denoted by $J_0$. We note that the numerical prefactors in Eqs. (\ref{eq:PC})--(\ref{eq:PC2}) differ slightly from those of Wong and Muthukumar \cite{wm}.

\begin{figure}
\centering
\resizebox{140mm}{!}{\includegraphics{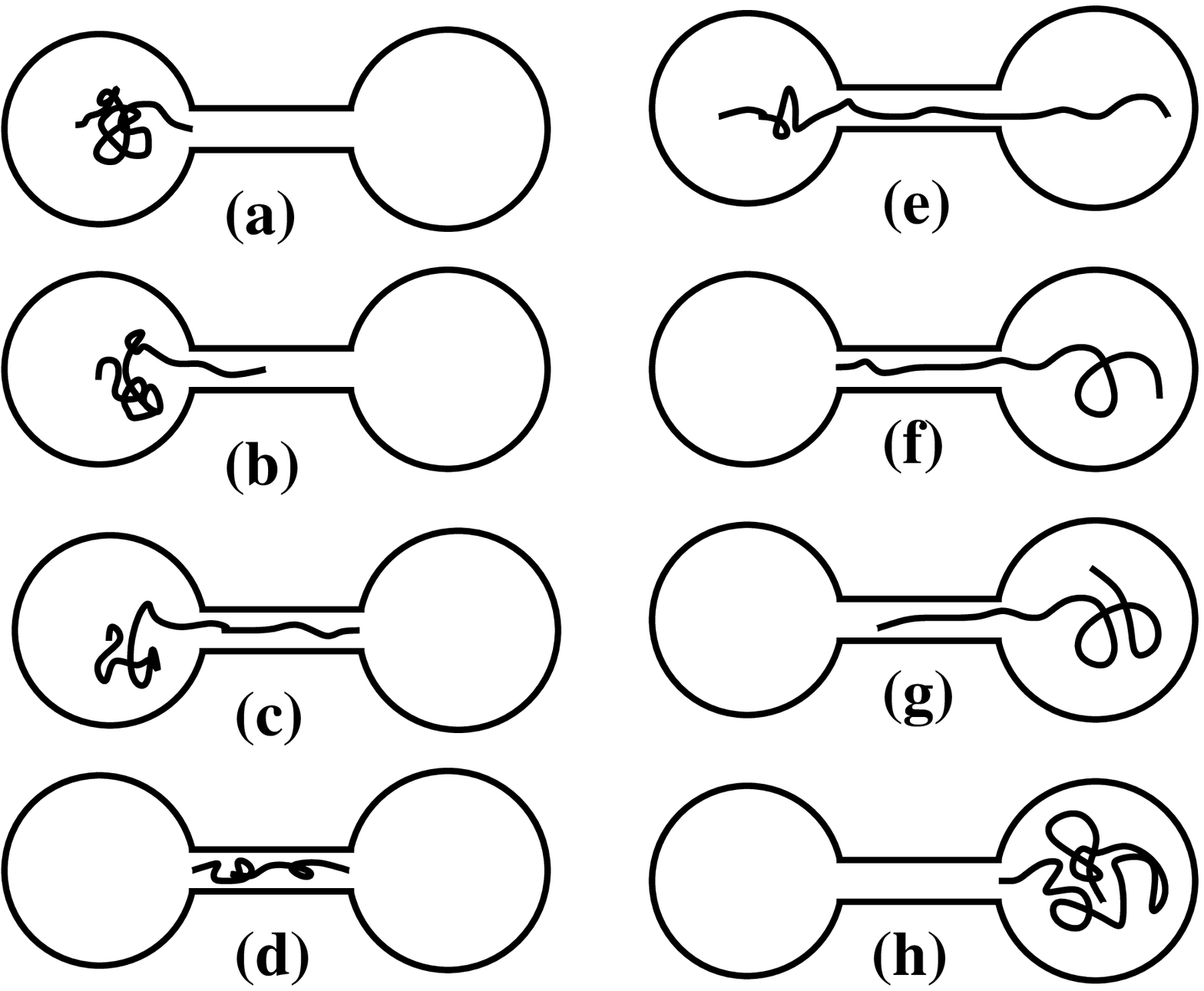} } 
\caption{Chain configurations during translocation from the donor to the receptor sphere through a cylindrical pore.}{ \label{fig:wmallpics}}
\end{figure}

Equations (\ref{eq:PS1}), (\ref{eq:PC1}) and (\ref{eq:PC2}), when appropriately combined, enable the computation of the free energy landscape for the translocation of a chain from the donor to the receptor sphere through a cylindrical pore, as illustrated by the sequence of chain configurations in Fig. \ref{fig:wmallpics}. As evident from Fig. \ref{fig:wmallpics}, we restrict our attention to situations where the chain length greatly exceeds the pore length, and disregard situations wherein the chain completely enters the cylinder and subsequently diffuses to the receptor sphere entrance. We will elaborate on this assumption in the following section. Following Wong and Muthukumar, we denote by $p$ and $q$ the number of chain segments in the cylinder and in the receptor sphere, respectively. We further denote the number of segments in the cylinder just before the start of translocation into the receptor sphere [cf. Fig \ref{fig:wmallpics}(c) and (d)] by $R_p$, with $M \leq R_p \leq N$. The radii of the donor and the receptor spheres are denoted, respectively, by $R_1$ and $R_2$. The free energies of the chain configurations depicted in Fig. \ref{fig:wmallpics} are \cite{wm} (in units of $k_BT$) 

\begin{equation}
F_{(a)}=-\ln \left[ P_{S_{1}}(R_{1},N)\pi a^{2}\right] \label{eq:Fa}
\end{equation}
\begin{equation}
F_{(b)}(p<R_p) = -\ln \left[ P_{S_{1}}(R_{1},N-p)P_{C_{1}}\left( p\right) %
\right] 
\end{equation}
\begin{equation}
F_{(c)}(p=R_p<N) =-\ln \left[ P_{S_{1}}(R_{1},N-p)P_{C_{2}}\left( p\right) %
\right] 
\end{equation}
\begin{equation}
F_{(d)}(p=R_p=N)=-\ln \left[ P_{C_{2}}\left(N\right) \right]
\end{equation}
\begin{equation}
F_{(e)}(q, R_p<N-1) = -\ln \left[ P_{S_{1}}(R_{1},N-R_p-q)P_{C_{2}}\left(R_p\right) P_{S_{1}}(R_{2},q)\right]
\end{equation}
\begin{equation}
F_{(f)}(q, R_p<N) = -\ln \left[ P_{C_{2}}\left( N-q\right) P_{S_{1}}(R_{2},q)\right]
\end{equation}
\begin{equation}
F_{(g)}(q)= -\ln \left[ P_{C_{1}}\left( N-q\right) P_{S_{1}}(R_{2},q)\right]  
\end{equation}
\begin{equation}
F_{(h)}=-\ln \left[ P_{S_{1}}(R_{2},N)\pi a^{2}\right] \label{eq:Fh}
\end{equation}

In the subsequent analysis, all tethering distances are set to $1/2$ (in units of $b$) for simplicity \cite{muthujcp03}.

\subsection{Translocation time}

The sequence of steps constituting the translocation process may be divided into two stages for the sake of convenience. The first stage, which begins with the insertion of the first segment of the chain into the cylinder from the donor sphere and lasts until the initiation of translocation into the receptor sphere, is comprised of either the sequence (a)$\rightarrow$(b)$\rightarrow$(c) for $R_p<N$ or the sequence (a)$\rightarrow$(b)$\rightarrow$(d) for $R_p=N$ in Fig. \ref{fig:wmallpics}. The transport of the chain into the receptor chamber occurs entirely within the second stage of translocation, which may be comprised of either the sequence (c)$\rightarrow$(e)$\rightarrow$(f)$\rightarrow$(g)$\rightarrow$(h) for $R_p<N $ [where step (e) must be omitted if $R_p=N-1$] or the alternate sequence (d)$\rightarrow$(g)$\rightarrow$(h) for $R_p=N$.

The probability density $W_p(t)$ of the number of segments $p$ contained within the cylinder at time $t$ during the first stage of translocation is governed by the equation \cite{vankampen}
\begin{equation}
\frac{\partial W_p}{\partial t}=\frac{\partial }{\partial p}\left[
\frac{\partial F}{\partial p} W_p + \frac{%
\partial W_p}{\partial p}\right] \label{eq:Wp}
\end{equation}%
where we have selected the unit of time to be the reciprocal of the diffusion constant $k_0$ (in the notation of Wong and Muthukumar), and $F$ denotes the free energy of the chain [cf. Eqs. (\ref{eq:Fa})--(\ref{eq:Fh})]. Consistent with prior theoretical studies, we decouple the diffusion of the chain end from within the donor sphere to the cylinder entrance from the process of chain migration into the cylinder once the leading chain end has located the pore. Therefore, we impose a reflecting (no flux) boundary condition at $p=1$:
\begin{equation}
- \left[ \frac{\partial F}{\partial p} W_p + \frac{\partial W_p}{\partial p}%
\right]_{p=1} =0 \label{eq:BCp1}
\end{equation}%
The number of segments $R_p$ in the cylinder at the end of the first stage of translocation must be at least $M$ and is at most $N$ [cf. Fig \ref{fig:wmallpics}(c) and (d)]. Moreover, the leading end of the chain may arrive at the cylinder--receptor sphere interface many times before finally being ``absorbed,'' thereby initiating the second stage of translocation. For this reason, we impose the following radiation boundary condition \cite{szabo} in place of the absorbing boundary condition at $p=R_p$:
\begin{equation}
- \left[ \frac{\partial F}{\partial p} W_p + \frac{\partial W_p}{\partial p}%
\right]_{p=R_p} = k_R W_{R_p}(t) \label{eq:BCpRp}  
\end{equation}%
where $k_R$ is a ``reaction'' rate constant, expressed in units of $k_0$. The radiation boundary condition reduces to the absorbing boundary condition in the limit $k_R \rightarrow \infty$, and to the reflecting boundary condition in the opposite limit $k_R \rightarrow 0$. The average passage time to reach $p=R_p$ starting from $p=1$ for a process governed by Eq. (\ref{eq:Wp}) and boundary conditions (\ref{eq:BCp1}) and (\ref{eq:BCpRp}) is given by the expression \cite{szabo}
\begin{multline}
\tau_1(R_p) = \int_{1}^{R_p}dy^{\prime }\exp \left[ F(p=y^{\prime })%
\right] \int_{1}^{y^{\prime }}dy^{\prime \prime }\exp \left[ -F(p=y^{\prime \prime })\right] \\
+ \frac{1}{k_{R}}\exp \left[ F(p=R_p)\right] \int_{1}^{R_p} dy^{\prime }\exp \left[ - F(p=y^{\prime })\right]
\label{eq:tau1Rp}
\end{multline}

We emphasize that the chain configurations illustrated in Fig. \ref{fig:wmallpics} are exhaustive only under the assumption $N \gg M$. We enforce this condition so as to avoid situations wherein the chain completely enters the pore and subsequently diffuses across the length of the pore to the receptor sphere entrance. The latter situation, which arises when $N \ll M$, necessitates the augmentation of the translocation time by the inclusion of a diffusion time in addition to the time taken by the chain to completely cross into the cylinder. We here restrict our analysis to the physically realistic situation wherein $N \gg M$.

Likewise, the probability density $W_q(t)$ of the number of segments $q$ present in the receptor sphere at time $t$ during the second stage of translocation is governed by the equation 
\begin{equation}
\frac{\partial W_q}{\partial t}=  \frac{\partial }{\partial q}\left[
\frac{\partial F}{\partial q} W_q + \frac{%
\partial W_q}{\partial q}\right] \label{eq:Wq}
\end{equation}%
subject to a reflecting boundary condition at $q=1$ and an absorbing boundary condition at $q=N$. The latter absorbing boundary condition is motivated by the fact that only successful translocation events are recorded, while the reflecting boundary condition at $q=1$ is consistent with the use of a radiation boundary condition at $p=R_p$, signifying the initiation of translocation into the receptor sphere. 
The corresponding mean first passage time is \cite{vankampen}
\begin{equation}
\tau_2(R_p) = \int_{1}^{N-1} dy^{\prime }\exp \left[  F(q=y^{\prime })%
\right] \int_{1}^{y^{\prime }}dy^{\prime \prime }\exp \left[ -
F(q=y^{\prime \prime })\right]  \label{eq:tau2Rp}
\end{equation}

For given $R_p$, Eqs. (\ref{eq:tau1Rp}) and (\ref{eq:tau2Rp}) yield the corresponding times of translocation, $\tau_1(R_p)$ and $\tau_2(R_p)$. We equate the average durations $\left\langle \tau_1 \right\rangle$ and $\left\langle \tau_2 \right\rangle$ of the first and second stages of translocation, respectively, to the average of  $\tau_1(R_p)$ and $\tau_2(R_p)$, respectively, computed with respect to the probability of configurations having $R_p$ segments in the cylinder for $R_p=M, M+1,...N$ [cf. Fig \ref{fig:wmallpics}(c) and (d)] as follows:
\begin{equation}
\left \langle \tau _{1} \right \rangle =\frac{\sum_{R_p=M}^{N-1}P_{S_{1}}(R_{1},N-R_p)P_{C_{2}}(R_p)\tau
_1(R_p)+P_{C_{2}}(N)\tau _1(R_p=N)}{%
\sum_{R_p=M}^{N-1}P_{S_{1}}(R_{1},N-R_p)P_{C_{2}}(R_p)+P_{C_{2}}(N)}  \label{eq:tau1wm}
\end{equation}
and
\begin{equation}
\left \langle \tau _{2} \right \rangle =\frac{\sum_{R_p=M}^{N-1}P_{S_{1}}(R_{1},N-R_p)P_{C_{2}}(R_p)\tau
_2(R_p)+P_{C_{2}}(N)\tau _2(R_p=N)}{%
\sum_{R_p=M}^{N-1}P_{S_{1}}(R_{1},N-R_p)P_{C_{2}}(R_p)+P_{C_{2}}(N)} \label{eq:tau2wm}
\end{equation}
The total time of translocation may now be defined as $ \left \langle \tau \right \rangle  = \left \langle \tau_1 \right \rangle  + \left \langle \tau_2 \right \rangle$.

The rate constants $k_R$, which are dependent on the value of $R_p$, remain to be specified. These are derived from the radiation boundary condition given by Eq. (\ref{eq:BCpRp}) in conjunction with the backward finite difference approximation
\begin{equation}
 \left.  \frac{\partial F}{\partial p}\right|_{p=R_p} =  
 \begin{cases}
 -\ln \left[P_{S_{1}}(R_{1},N-R_p)P_{C_{2}}(R_p)\right] +\ln \left[
P_{S_{1}}(R_{1},N-R_p+1)P_{C_{1}}(R_p-1)\right], \ R_p<N \\
-\ln \left[P_{C_{2}}(N)\right] +\ln \left[ P_{S_{1}}(R_{1},1)P_{C_{1}}(N-1)\right], \ R_p=N
\end{cases} \label{eq:derFp}
\end{equation}
We further replace the term $W_p(t)$ appearing in Eq. (\ref{eq:BCpRp}) with the probabilities of the corresponding configurations:
\begin{equation}
W_p(p=R_p) \sim
\begin{cases} 
P_{S_{1}}(R_{1},N-R_p)P_{C_{2}}(R_p), \ R_p<N \\
P_{C_{2}}(N), \ R_p=N
\end{cases}
\end{equation}%
and the backward finite difference approximation yields
\begin{equation}
\left. \frac{\partial W_p}{\partial p}\right|_{p=R_p} \sim
\begin{cases}
P_{S_{1}}(R_{1},N-R_p)P_{C_{2}}(R_p)-P_{S_{1}}(R_{1},N-R_p+1)P_{C_{1}}(R_p-1), \ R_p<N \\
P_{C_{2}}(N)-P_{S_{1}}(R_{1},1)P_{C_{1}}(N-1), \ R_p=N
\end{cases} \label{eq:derwp}
\end{equation}
We note that the simultaneous derivation of $F$ and $W_p$ from the probabilities of the chain configurations would result in a vanishing $k_R$ for a continuous process [cf. Eq. (\ref{eq:BCpRp})]. However, in the present case of discrete chain transport in the absence of external forces or fields, the diffusive contribution to the probability flux in Eq. (\ref{eq:BCpRp}) dominates over the convective contribution, owing to the large change in the probabilities of the corresponding chain configurations when $p$ changes from $R_p-1$ to $R_p$. In our calculations, values of $k_R$ obtained were in the range of $O(1)$--$O(100)$.

Use of Eqs. (\ref{eq:tau1wm}) and (\ref{eq:tau2wm}), following the numerical integration of (\ref{eq:tau1Rp}) and (\ref{eq:tau2Rp}) in combination with (\ref{eq:Fa})--(\ref{eq:Fh}), (\ref{eq:BCpRp}) and (\ref{eq:derFp})--(\ref{eq:derwp}), finally yields the desired translocation times.

\subsection{Results}

\begin{figure}
\centering
\resizebox{70mm}{!}{\includegraphics{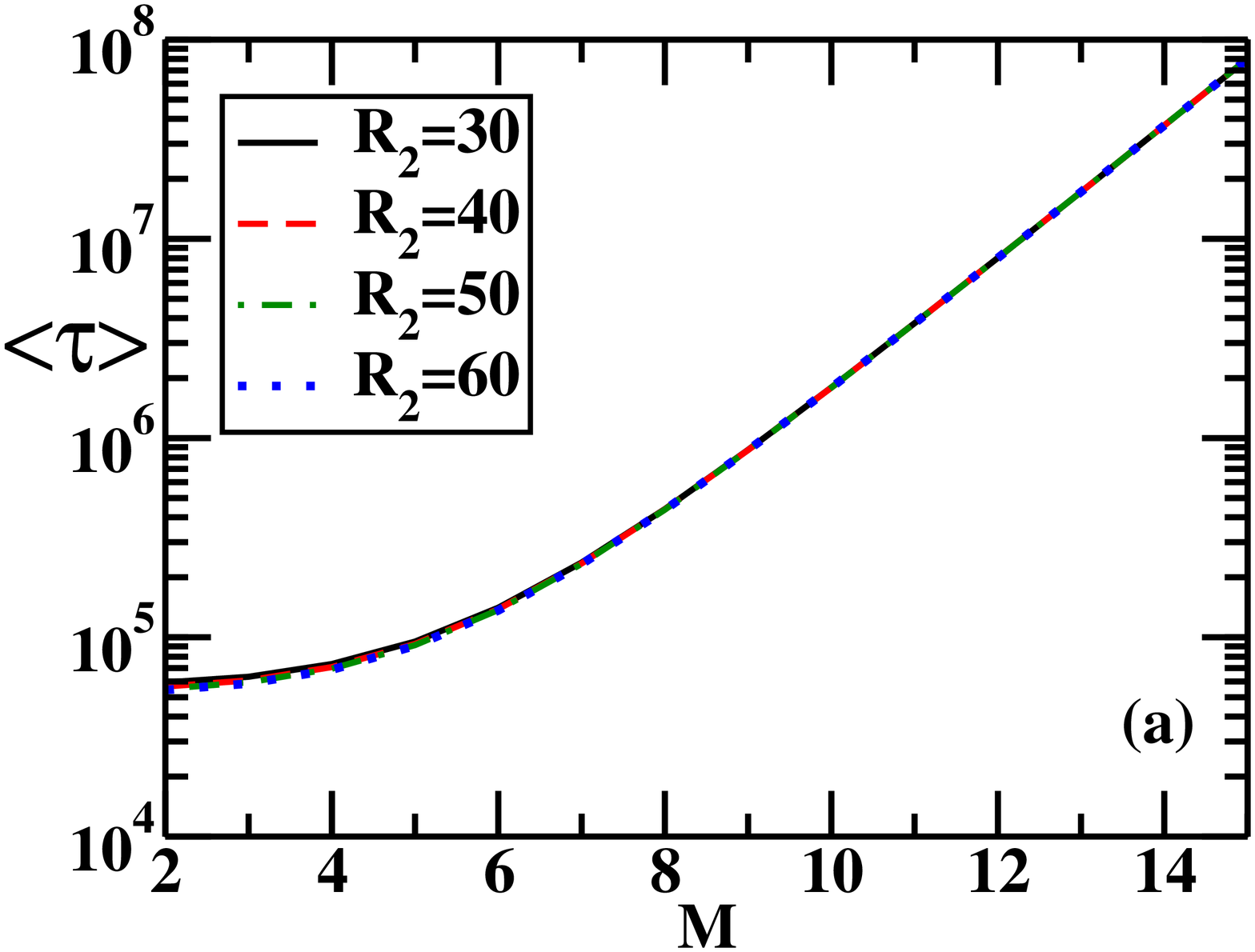} } 
\vskip 2.25em
\centering
\resizebox{70mm}{!}{\includegraphics{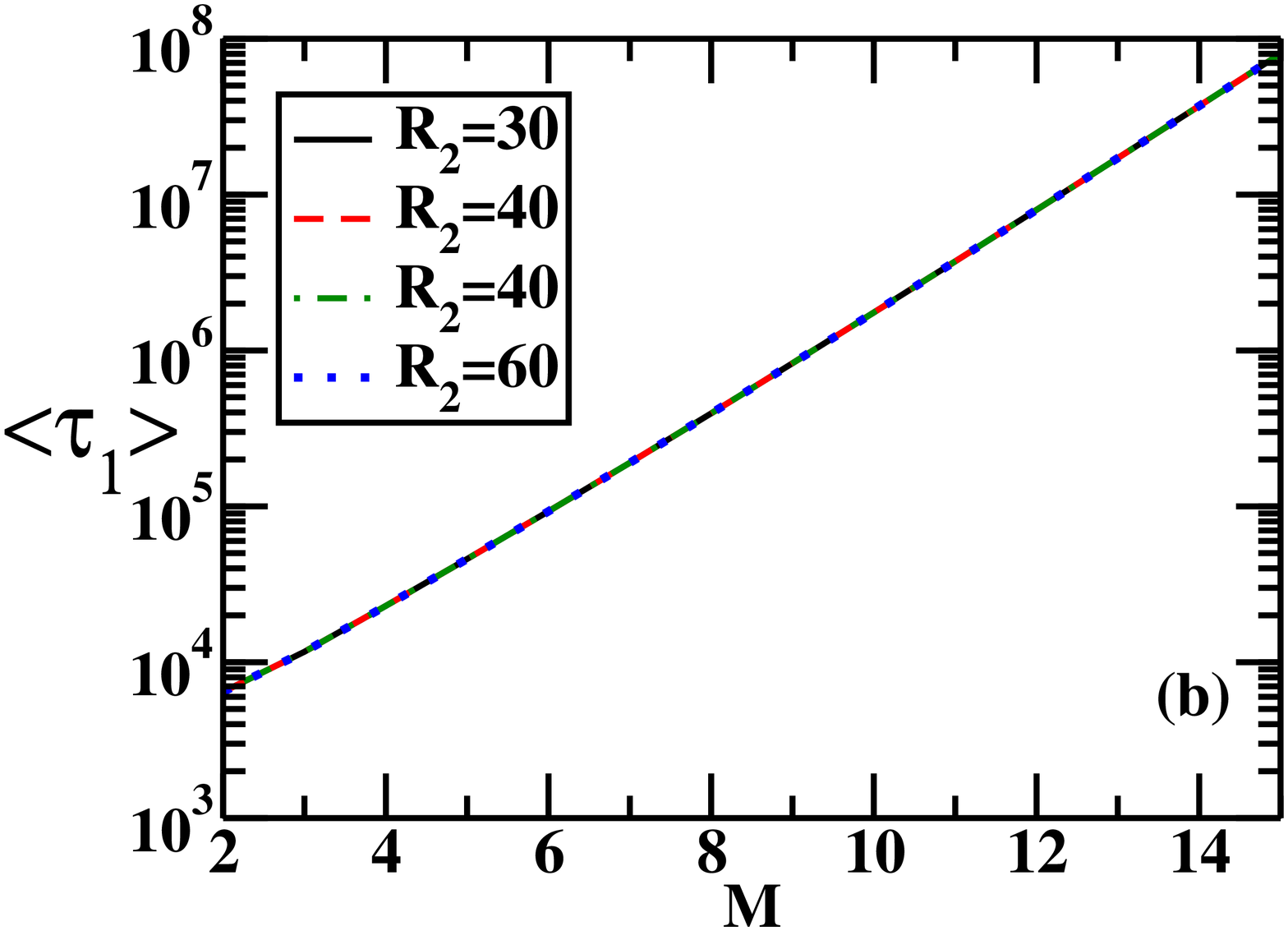} } 
\vskip 2.25em
\centering
\resizebox{70mm}{!}{\includegraphics{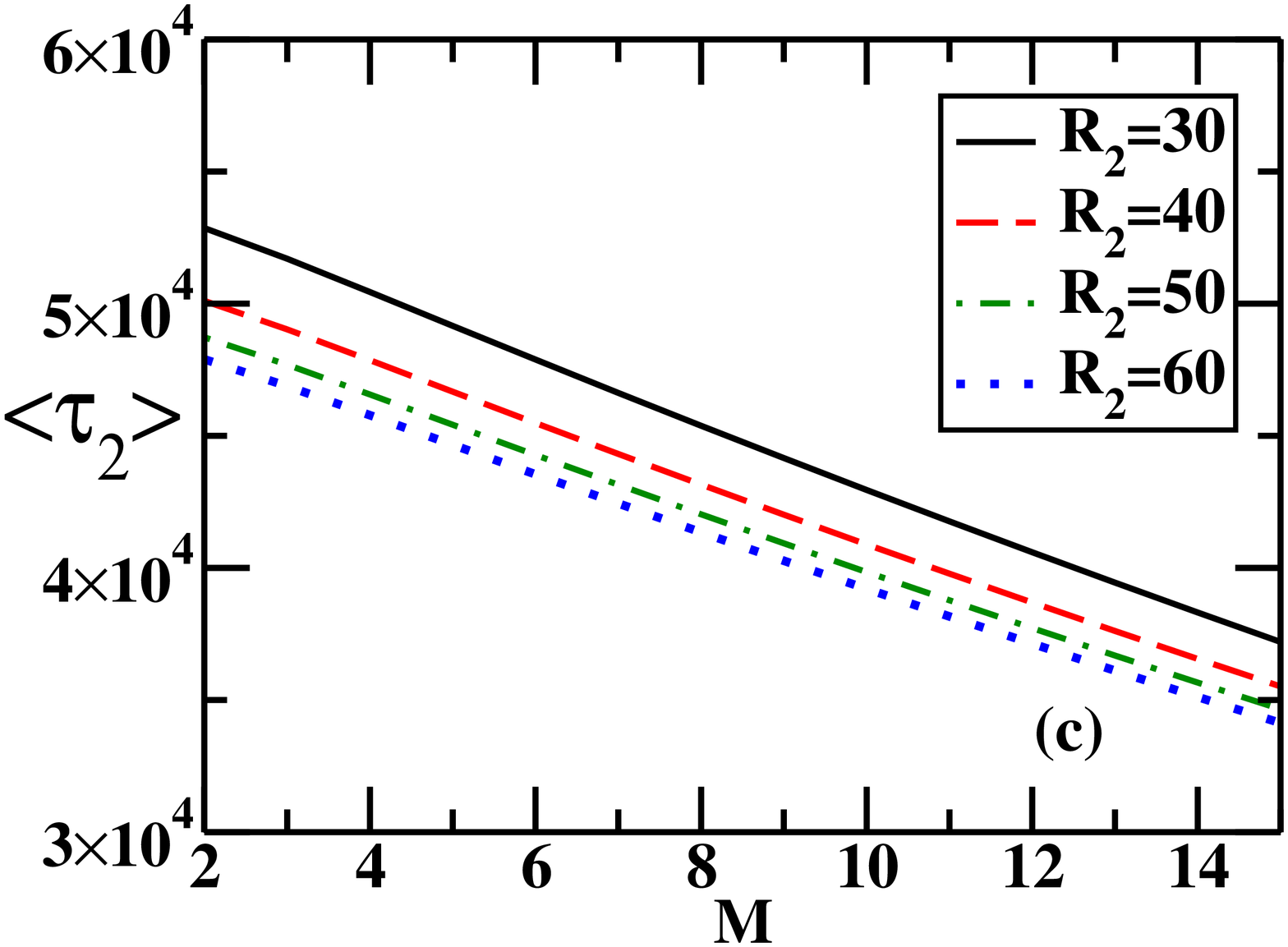} }
\caption{(Color online) (a) Total average translocation time, (b) average time taken for the completion of the first stage of translocation, and (c) average time taken for the completion of the second stage of translocation from the donor to the receptor sphere through a cylindrical pore as a function of cylinder length $M$ for several values of $R_2$ with $R_1=30$ and $a=3$ for a chain possessing $N=300$ segments.}{ \label{fig:varR2} }
\end{figure}

\begin{figure}
\centering
\resizebox{70mm}{!}{\includegraphics{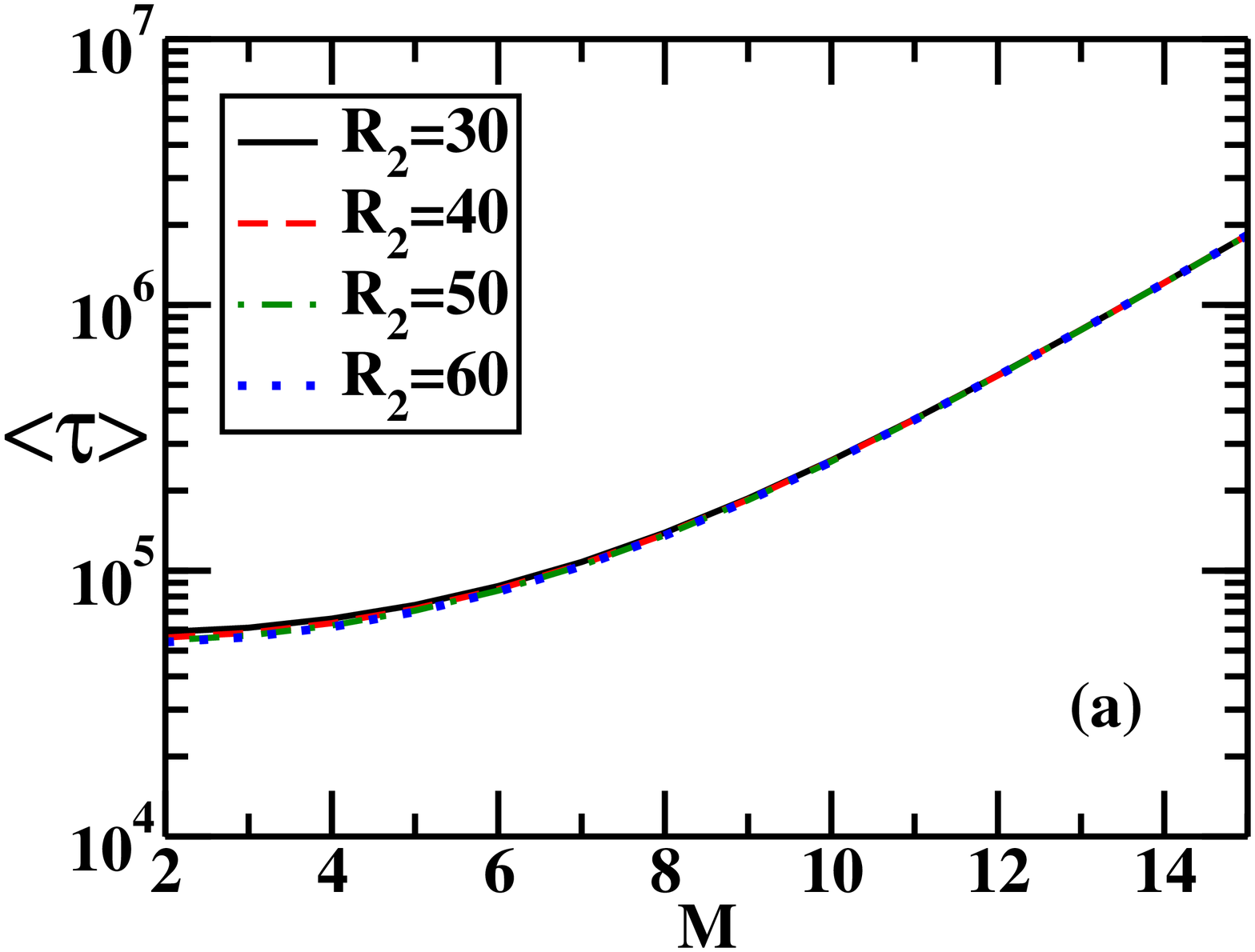} } 
\vskip 2.25em
\centering
\resizebox{70mm}{!}{\includegraphics{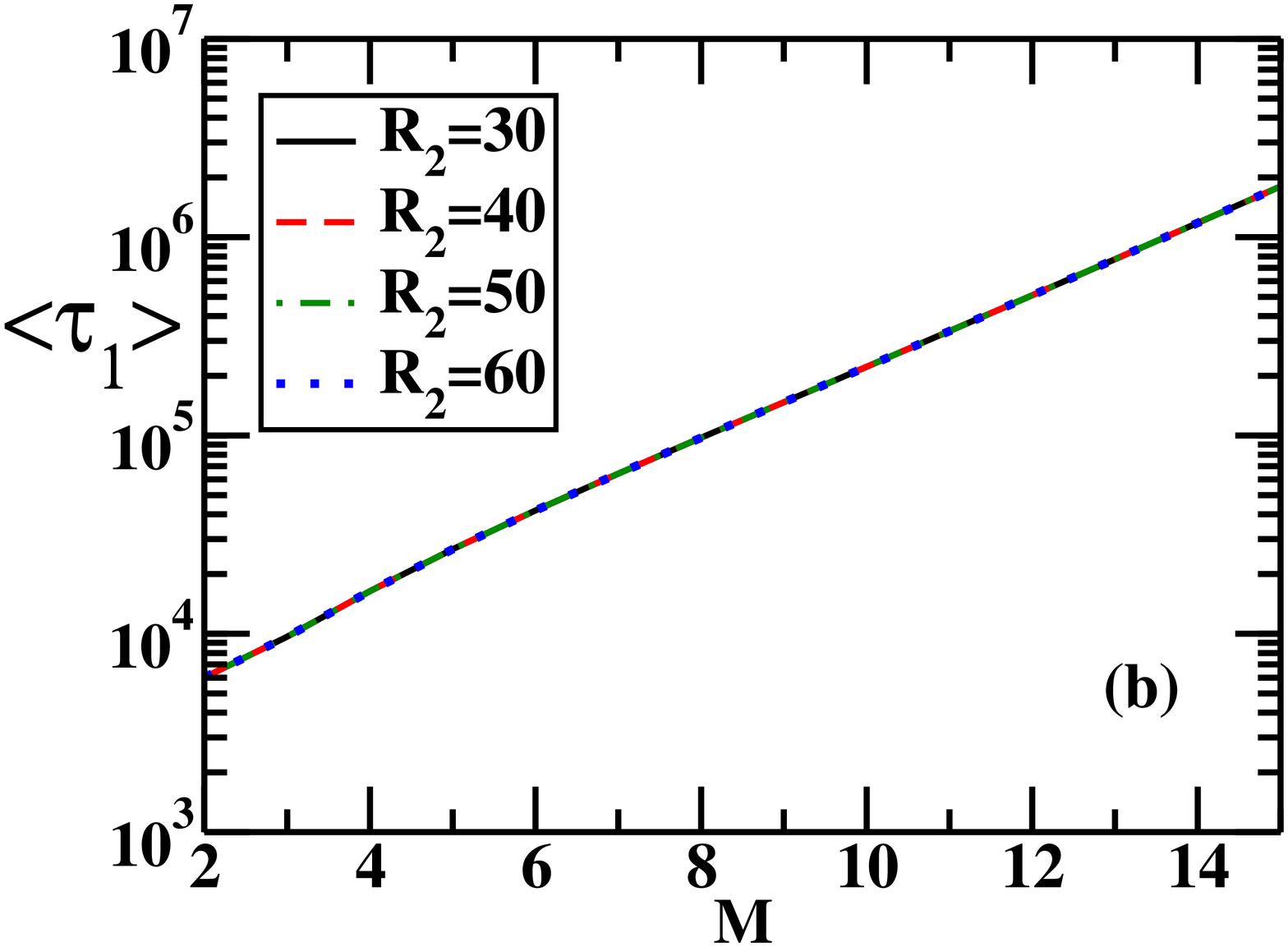} } 
\vskip 2.25em
\centering
\resizebox{70mm}{!}{\includegraphics{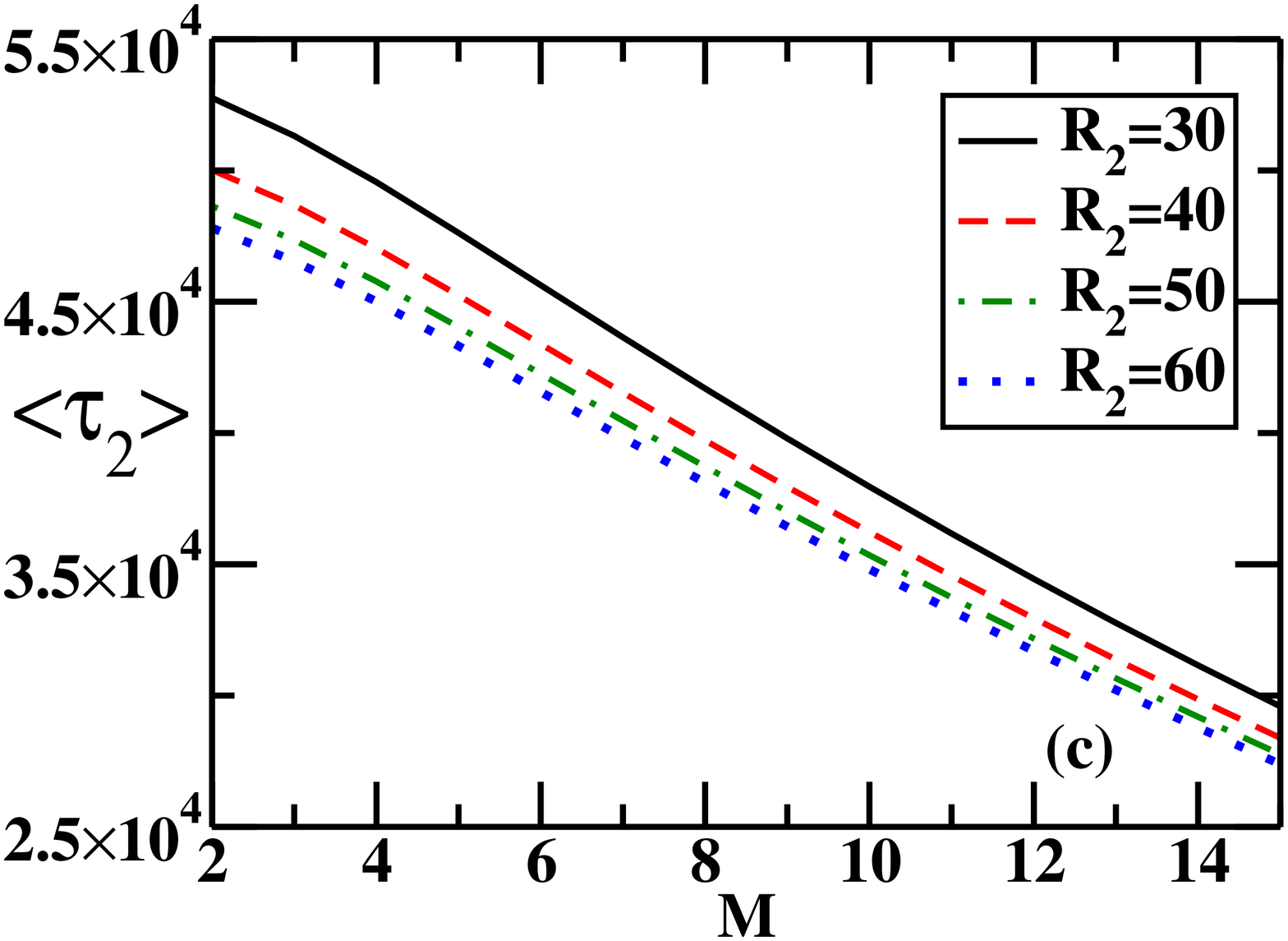} }
\caption{(Color online) (a) Total average translocation time, (b) average time taken for the completion of the first stage of translocation, and (c) average time taken for the completion of the second stage of translocation from the donor to the receptor sphere through a cylindrical pore as a function of cylinder length $M$ for several values of $R_2$ with $R_1=30$ and $a=5$ for a chain possessing $N=300$ segments.} {\label{fig:a5varR2} }
\end{figure}

\begin{figure}
\centering
\resizebox{70mm}{!}{\includegraphics{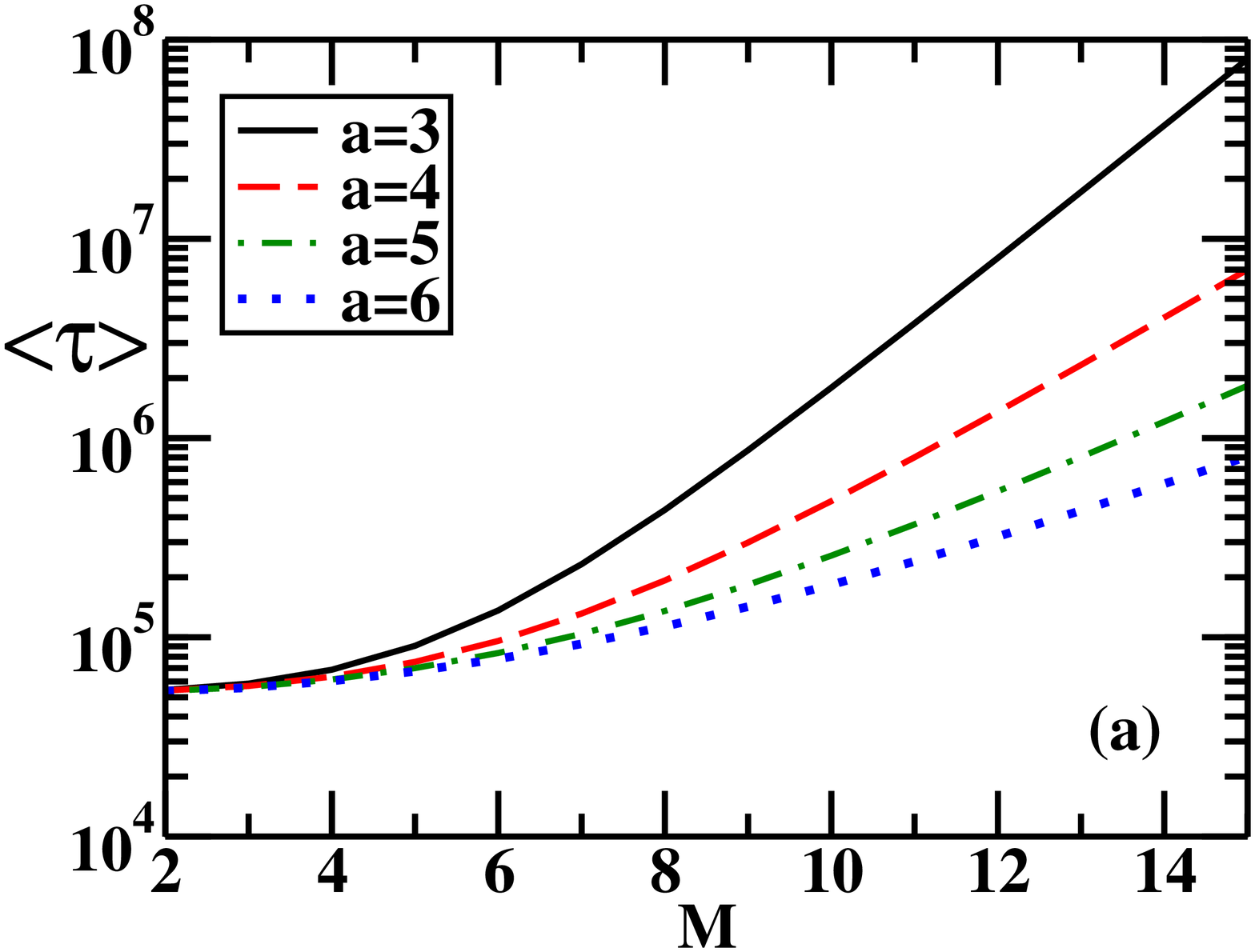} } 
\vskip 2.25em
\centering
\resizebox{70mm}{!}{\includegraphics{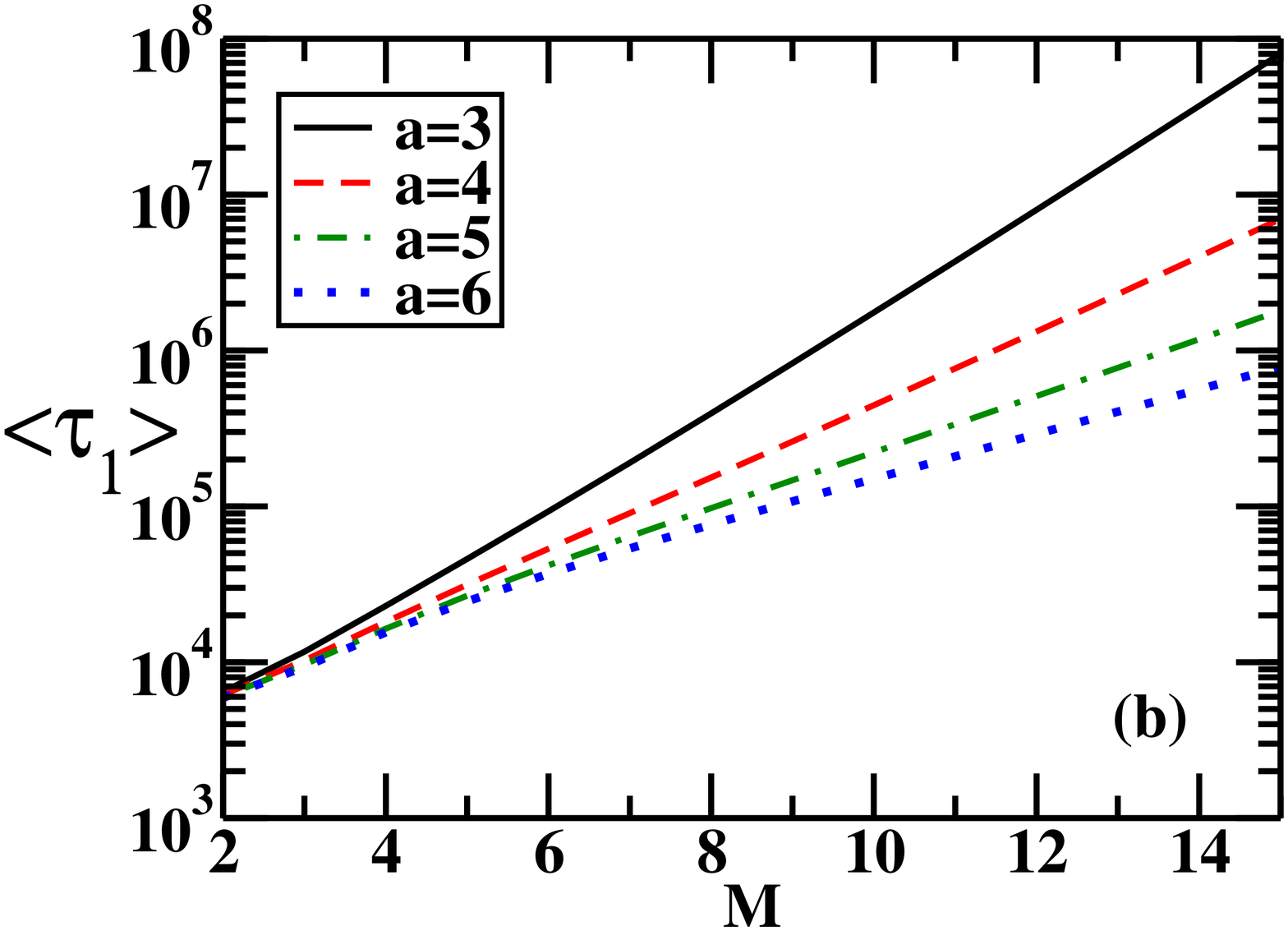} } 
\vskip 2.25em
\centering
\resizebox{70mm}{!}{\includegraphics{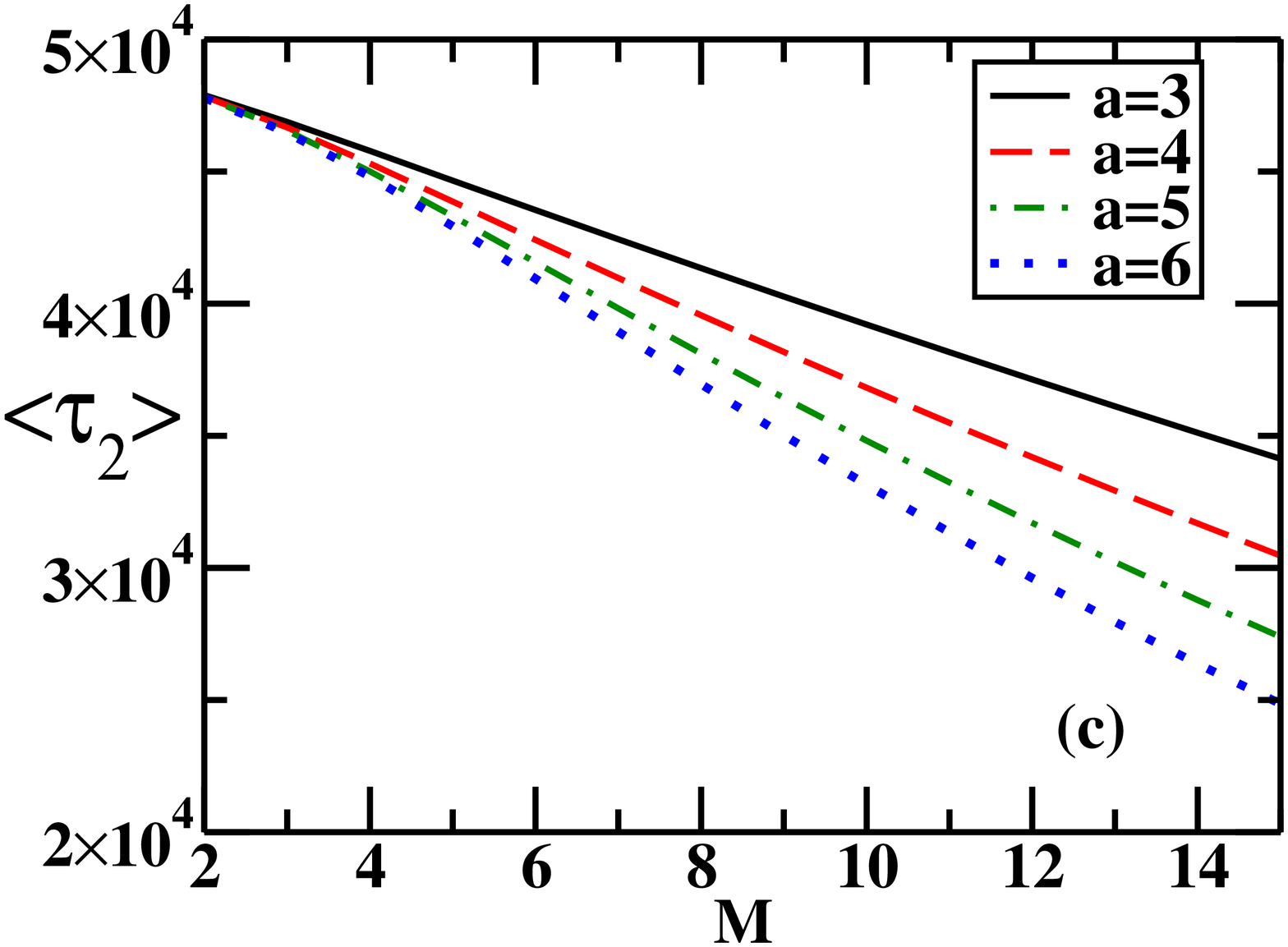} }
\caption{(Color online) (a) Total average translocation time, (b) average time taken for the completion of the first stage of translocation, and (c) average time taken for the completion of the second stage of translocation from the donor to the receptor sphere through a cylindrical pore as a function of cylinder length $M$ for several values of $a$ with $R_1=30$ and $R_2=60$ for a chain possessing $N=300$ segments.} {\label{fig:vara} }
\end{figure}

Our results for a chain of $N=300$ segments are illustrated in Figs. \ref{fig:varR2}--\ref{fig:vara}, where length and time have been expressed in units of $b$ and $k_0^{-1}$, respectively.  Figures \ref{fig:varR2} and \ref{fig:a5varR2} depict translocation times as a function of $M$ for several values of $R_2$ with $R_1=30$, and $a=3$ and $a=5$, respectively. It is evident that, contrary to the observations of Wong and Muthukumar \cite{wm}, the total translocation time grows monotonically with respect to increasing cylinder length. The results suggest that backward motions at the cylinder--receptor sphere interface do, in fact, significantly contribute to the duration of the first stage of translocation. The chain incurs an entropic penalty upon moving into the cylinder from the donor sphere, whereas there is an increase in entropy during translocation into the receptor sphere. Hence, the total translocation time $\left\langle \tau \right\rangle$ is dominated by the contribution from the first stage of translocation and, consequently, reflects the increase in $\left\langle \tau_1 \right\rangle$ with increasing $M$.

The duration of the second stage of translocation decreases as $M$ is increased, owing to the fact that chain configurations in which a larger number of segments are present in the cylinder have a higher free energy than those in which a larger number of segments are present in the donor sphere. As a result, translocation into the receptor sphere leads to a greater free energy drop as $M$ is increased. This observation was made earlier by Wong and Muthukumar. However, the second stage of translocation makes a far smaller contribution to the total translocation time relative to the first stage.

Figure \ref{fig:vara} illustrates the translocation times as a function of $M$ for several values of $a$, with $R_1=30$ and $R_2=60$. As expected, an increase in $a$ causes a decrease in $\left\langle \tau \right\rangle$ and $\left\langle \tau_1 \right\rangle$, owing to the concomitant lowering of the entropic barrier to translocation into the cylinder. Moreover, configurations in which a larger number of chain segments reside in the cylinder become more favorable as $a$ is increased relative to those in which the majority of chain segments lie in the donor sphere. Since the former configurations have a higher free energy than the latter, the driving force for translocation into the receptor sphere is greater in the former case. Consequently, $\left\langle \tau_2 \right\rangle$ decreases with increase in $a$.

\section{\label{sec:ah} Two spheres connected by a composite two-cylinder pore}

In this section, we consider the translocation of a chain from a donor sphere of radius $R_1$ to a receptor sphere of radius $R_2$, connected by a composite two-cylinder pore with radii $a_1$ and $a_2$ and lengths $M_1$ and $M_2$, where the subscripts $1$ and $2$ refer to the outer and the inner cylinder, respectively. We now divide the translocation process into three stages, namely, (1) the transport of the chain into the outer cylinder until the leading chain end enters the inner cylinder, (2) the transport of the chain into the inner cylinder until its leading end enters the receptor sphere, and (3) the complete migration of the chain into the receptor sphere. We denote by $l$, $m$ and $n$, the number of segments in the outer cylinder, inner cylinder and receptor sphere, respectively, and by $R_l$ and $R_m$ the maximum number of chain segments in the outer cylinder and inner cylinder, respectively, just prior to the start of the succeeding stage. Again, we have the conditions that $M_1 \leq R_l \leq N$, and $M_2 \leq R_m \leq N$.

\begin{figure}
\centering
\resizebox{135mm}{!}{\includegraphics{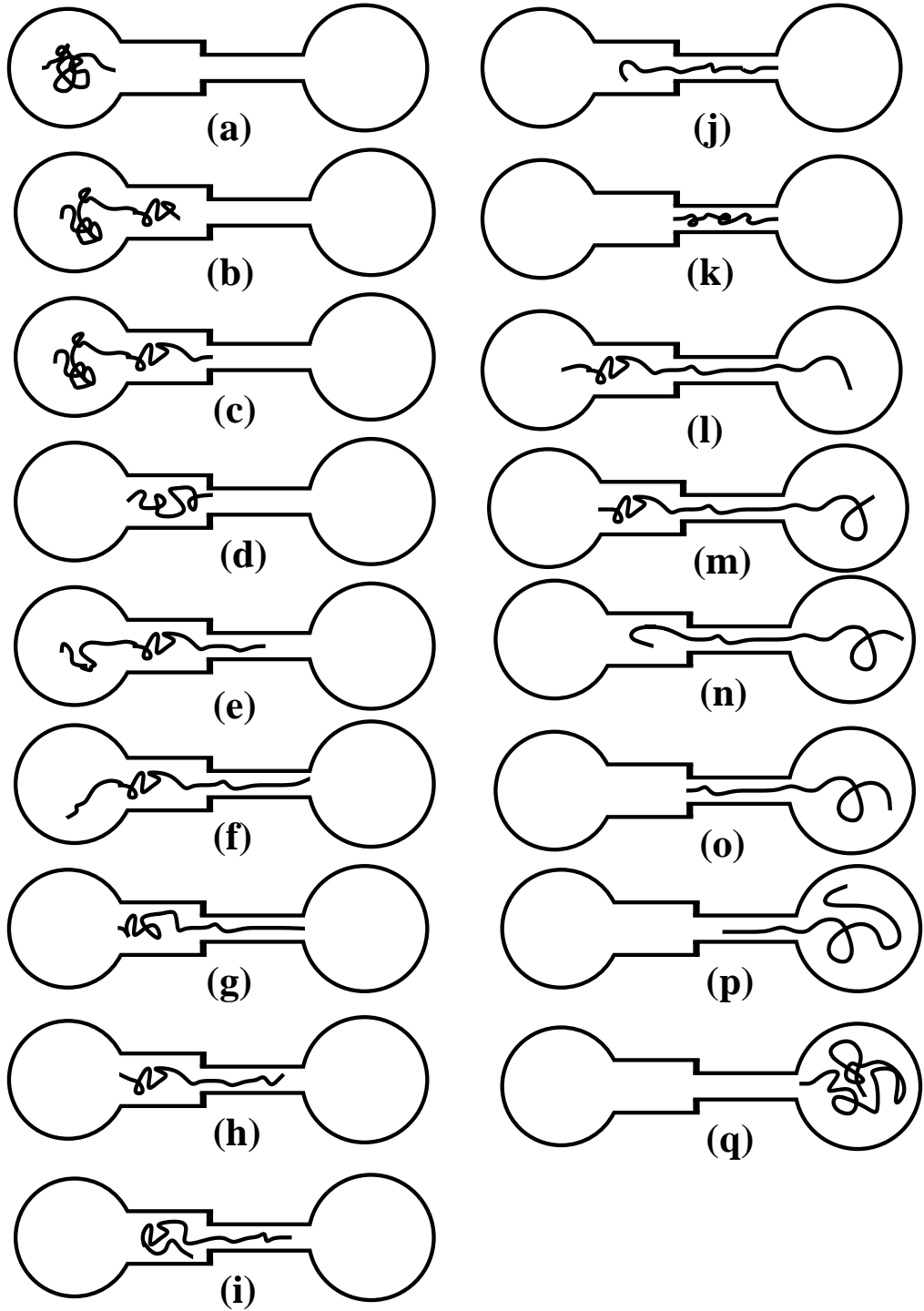} } 
\caption{Chain configurations during translocation from the donor to the receptor sphere through an $\alpha$-hemolysin membrane channel represented as a composite two-cylinder pore. For the purpose of illustration, we have selected the case $a_1>a_2$.}{ \label{fig:ahallpics} }
\end{figure}

Under the assumption that the chain length greatly exceeds the lengths of the cylinders, Fig. \ref{fig:ahallpics} illustrates all possible chain configurations during translocation. The first stage of translocation comprises either the sequence (a)$\rightarrow$(b)$\rightarrow$(c) for $R_l<N$, or the sequence (a)$\rightarrow$(b)$\rightarrow$(d) for $R_l=N$ (cf. Fig \ref{fig:ahallpics}). Similarly, the second stage of translocation is identified with one of the following sequences: (c)$\rightarrow$(e)$\rightarrow$(f) for $R_l+R_m<N$; (c)$\rightarrow$(e)$\rightarrow$(g) for $R_l+R_m=N$; (c)$\rightarrow$(e)$\rightarrow$(h)$\rightarrow$(i)$\rightarrow$(j) for $R_l<N$, $R_m<N$ and $R_l+R_m>N$; (c)$\rightarrow$(e)$\rightarrow$(h)$\rightarrow$(i)$\rightarrow$(k) for $R_l<N$, $R_m=N$; (d)$\rightarrow$(i)$\rightarrow$(j) for $R_l=N$ and $R_m<N$; and (d)$\rightarrow$(i)$\rightarrow$(k) for $R_l=N$ and $R_m=N$. Finally, the third stage of translocation is described by one of the following: (f)$\rightarrow$(l)$\rightarrow$(m)$\rightarrow$(n)$\rightarrow$(o)$\rightarrow$(p)$\rightarrow$(q) for $R_l+R_m<N$ (where the intermediate step (l) must be omitted if $R_l+R_m=N-1$); 
(g)$\rightarrow$(n)$\rightarrow$(o)$\rightarrow$(p)$\rightarrow$(q) for $R_l+R_m = N$; 
(j)$\rightarrow$(n)$\rightarrow$(o)$\rightarrow$(p)$\rightarrow$(q) for $R_m<N$ and $R_l+R_m > N$; and
(k)$\rightarrow$(p)$\rightarrow$(q) for $R_m = N$. The free energies of these chain configurations are listed in the Appendix.

The probability densities $W_l,\ W_m\ \text{and}\ W_n$ of the translocation coordinates $l$, $m$ and $n$, respectively, are governed by the equation
\begin{equation}
\frac{\partial W_x}{\partial t}= \frac{\partial }{\partial x}\left[
 \frac{\partial F}{\partial x} W_x + \frac{%
\partial W_x}{\partial x}\right] \label{eq:Wx}
\end{equation}%
where $x=l,\ m\ \text{or} \ n$, and $F$ denotes the free energy of the chain. We again impose a reflecting boundary condition at $l=1$ and an absorbing boundary condition at $n=N$, corresponding to experimental measurements wherein only successful translocation events are recorded. Since the leading chain end may visit the interface between the inner cylinder and the outer cylinder or the outer cylinder and the receptor sphere many times before being ``absorbed'' at the boundary, we impose the radiation boundary condition at $l=R_l$ and $m=R_m$, given by the expressions
\begin{equation}
- \left[ \frac{\partial F}{\partial
l}W_l+\frac{\partial W_l}{\partial l}\right] _{l=R_{l}} = k_1 W_l(l=R_l)  \label{eq:k1}
\end{equation}
and 
\begin{equation}
- \left[ \frac{\partial F  }{\partial
m}W_m+\frac{\partial W_m}{\partial m}\right] _{m=R_{m}} = k_2 W_m(m=R_m)  \label{eq:k2}
\end{equation}
where $k_1$ and $k_2$ are rate constants (in units of $k_0$), determined following the procedure described in the Appendix. In addition, we impose reflecting boundary conditions at $m=1$ and $n=1$, consistent with the above radiation boundary conditions.

The corresponding average passage times are yielded by the expressions
\begin{multline}
 \tau _{1}(R_{l})=\int_{1}^{R_{l}}dy^{\prime }\exp \left[
F(l=y^{\prime })\right] \int_{1}^{y^{\prime }}dy^{\prime \prime }\exp \left[
- F(l=y^{\prime \prime })\right] \\
+ \frac{1}{k_{1}}\exp \left[ 
F(l=R_{l})\right] \int_{1}^{R_{l}}dy^{\prime }\exp \left[ - F(l=y^{\prime })%
\right]   
\end{multline}
\begin{multline}
 \tau _{2}(R_{l},R_{m})= \int_{1}^{R_{m}}dy^{\prime }\exp \left[
F(m=y^{\prime })\right] \int_{1}^{y^{\prime }}dy^{\prime \prime }\exp %
\left[ - F(m=y^{\prime \prime })\right] \\
+ \frac{1}{k_{2}}\exp \left[ 
F(m=R_{m})\right] \int_{1}^{R_{m}}dy^{\prime }\exp \left[ - F(m=y^{\prime })%
\right]
\end{multline}
and
\begin{equation}
\tau _{3}\left( R_{l},R_{m}\right) = \int_{1}^{N-1}dy^{\prime
}\exp \left[  F(n=y^{\prime })\right] \int_{1}^{y^{\prime }}dy^{\prime
\prime }\exp \left[ - F(n=y^{\prime \prime })\right]
\end{equation}
We finally obtain the average translocation times from the expressions
\begin{equation}
\left \langle \tau _{1} \right\rangle =\frac{\sum_{R_{l}=M_{1}}^{N}P(R_l) \tau _{1}\left( R_{l}\right)}
{\sum_{R_{l}=M_{1}}^{N} P(R_l)}  \label{eq:tau1ah}
\end{equation}
\begin{equation}
\left \langle \tau _{2} \right \rangle =\frac{\sum_{R_{l}=M_{1}}^{N} \sum_{R_{m}=M_{2}}^{N} P \left(
R_{l}\right)P(R_{m}|R_{l})\tau _{2}\left(
R_{l},R_{m}\right) }{\sum_{R_{l}=M_{1}}^{N} \sum_{R_{m}=M_{2}}^{N} P\left( R_{l}\right) P(R_{m}|R_{l})}  \label{eq:tau2ah}
\end{equation}
and
\begin{equation}
\left \langle \tau _{3} \right \rangle =\frac{\sum_{R_{l}=M_{1}}^{N}%
\sum_{R_{m}=M_{2}}^{N}P \left(R_{l}\right)P(R_{m}|R_{l}) \tau _{3}\left(
R_{l},R_{m}\right) }{\sum_{R_{l}=M_{1}}^{N}\sum_{R_{m}=M_{2}}^{N}P \left(
R_{l}\right)P(R_{m}|R_{l}) } \label{eq:tau3ah}
\end{equation}
where $P(R_l)$ and $P(R_m|R_l)$ denote the probabilities of configurations with a maximum of $R_l$ segments in the outer cylinder, and a maximum of $R_m$ segments in the inner cylinder given the value of $R_l$, respectively.

Equations (\ref{eq:k1})--(\ref{eq:tau3ah}) enable the computation of the translocation times for each of the three stages, whereby we obtain the total translocation time $\left\langle \tau \right\rangle$ from the sum $\left\langle \tau_1 \right \rangle+ \left \langle \tau_2 \right\rangle + \left\langle \tau_3 \right\rangle$. The details of the calculation are analogous to those presented in Sec. \ref{sec:cylinder}, and are relegated to the Appendix.

\begin{figure}
\centering
\resizebox{83mm}{!}{\includegraphics{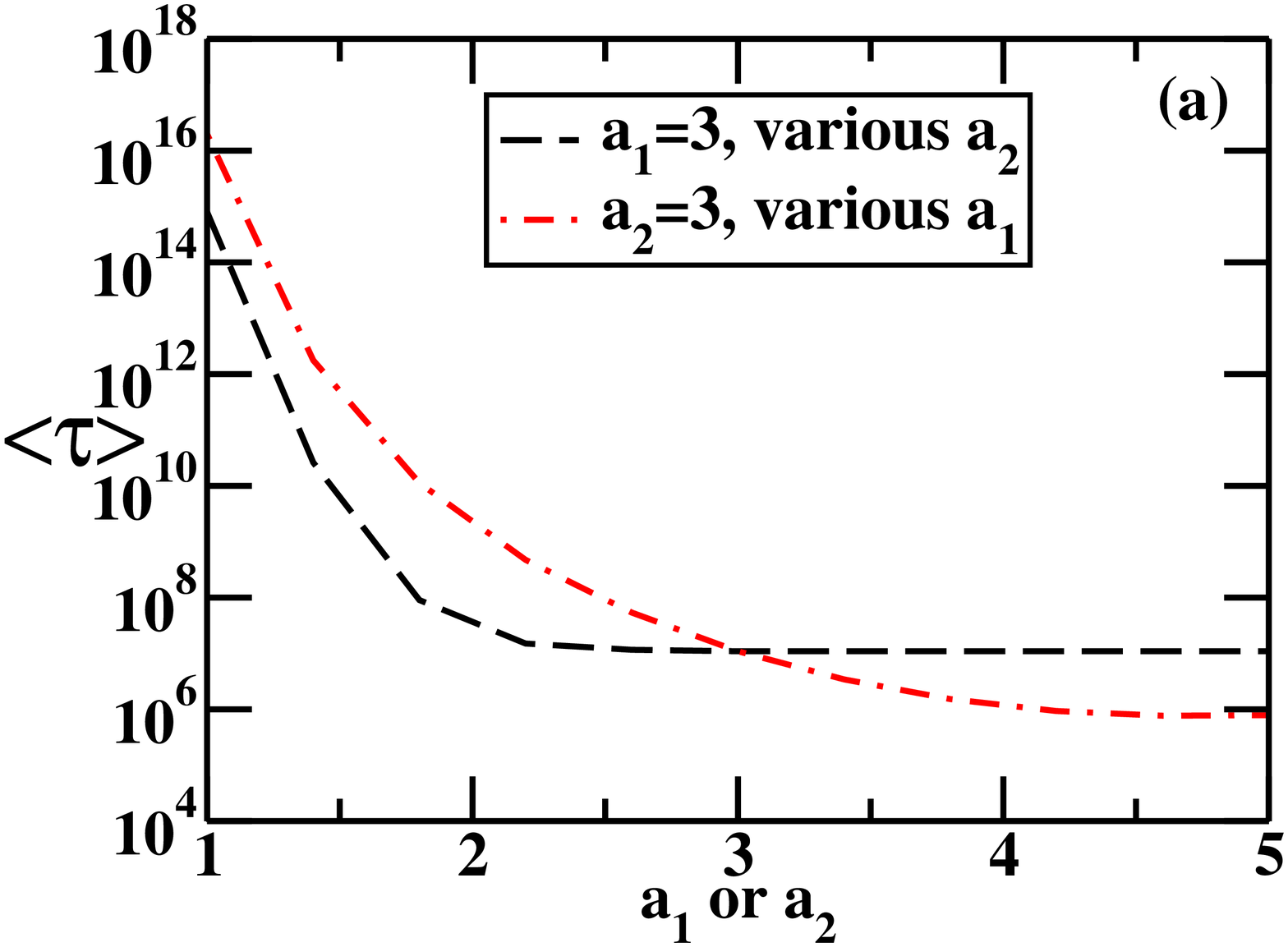} } 
\vskip 2.25em
\centering
\resizebox{83mm}{!}{\includegraphics{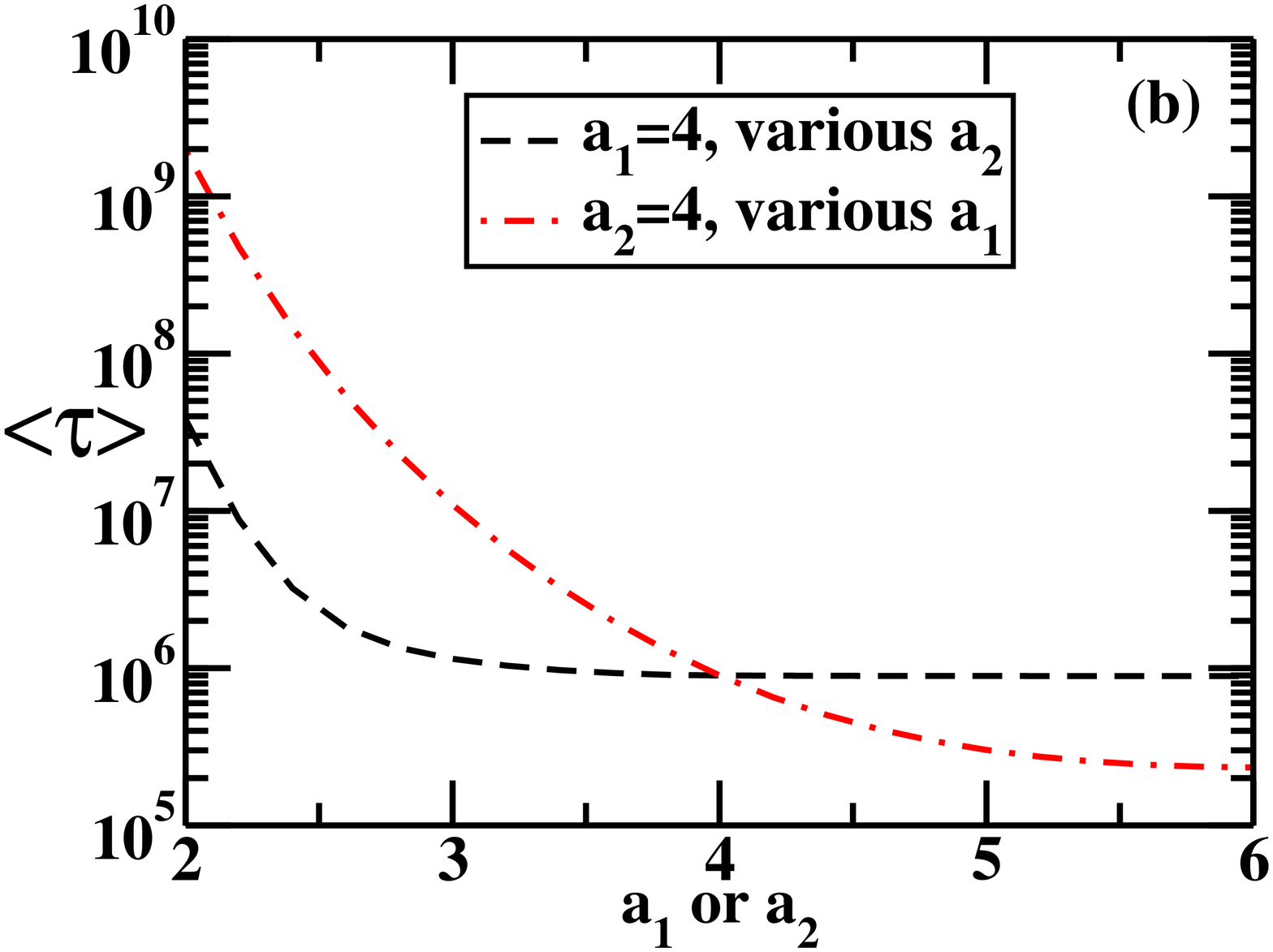} } 
\caption{(Color online) Average time of translocation from the donor to the receptor sphere through an $\alpha$-hemolysin channel with (a) an outer cylinder radius of $a_1=3$ for several values of the inner cylinder radius $a_2$ (dashed line) and vice versa (dotted-dashed line), and (b) an outer cylinder radius of $a_1=4$ for several values of the inner cylinder radius $a_2$ (dashed line) and vice versa (dotted-dashed line), and with $R_1=R_2=30$ and $M_1=M_2=15$, for a chain possessing $N=100$ segments.}{\label{fig:aha3a4tau} }
\end{figure}

\begin{figure}
\centering
\resizebox{70mm}{!}{\includegraphics{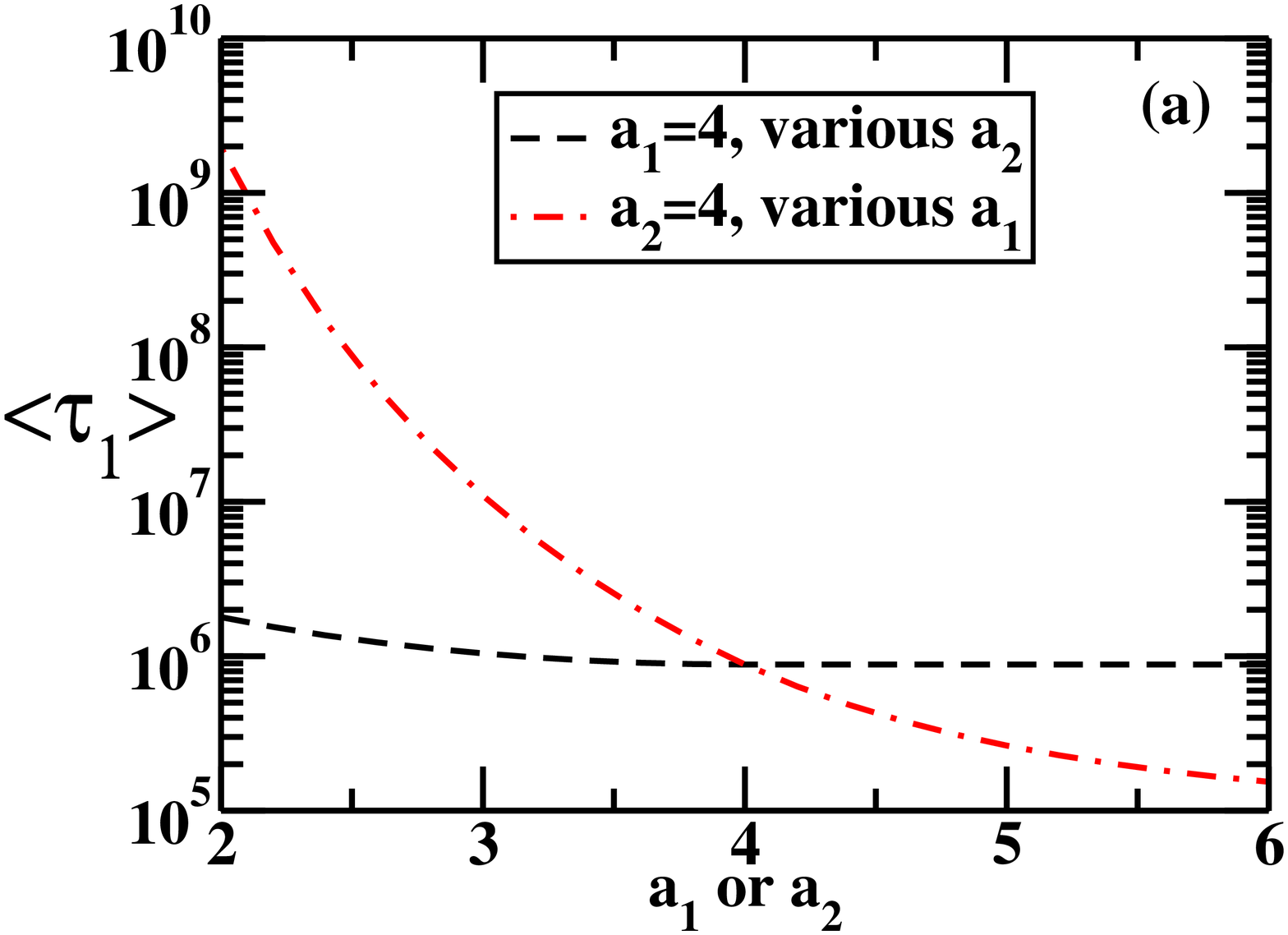} } 
\vskip 2.25em
\centering
\resizebox{70mm}{!}{\includegraphics{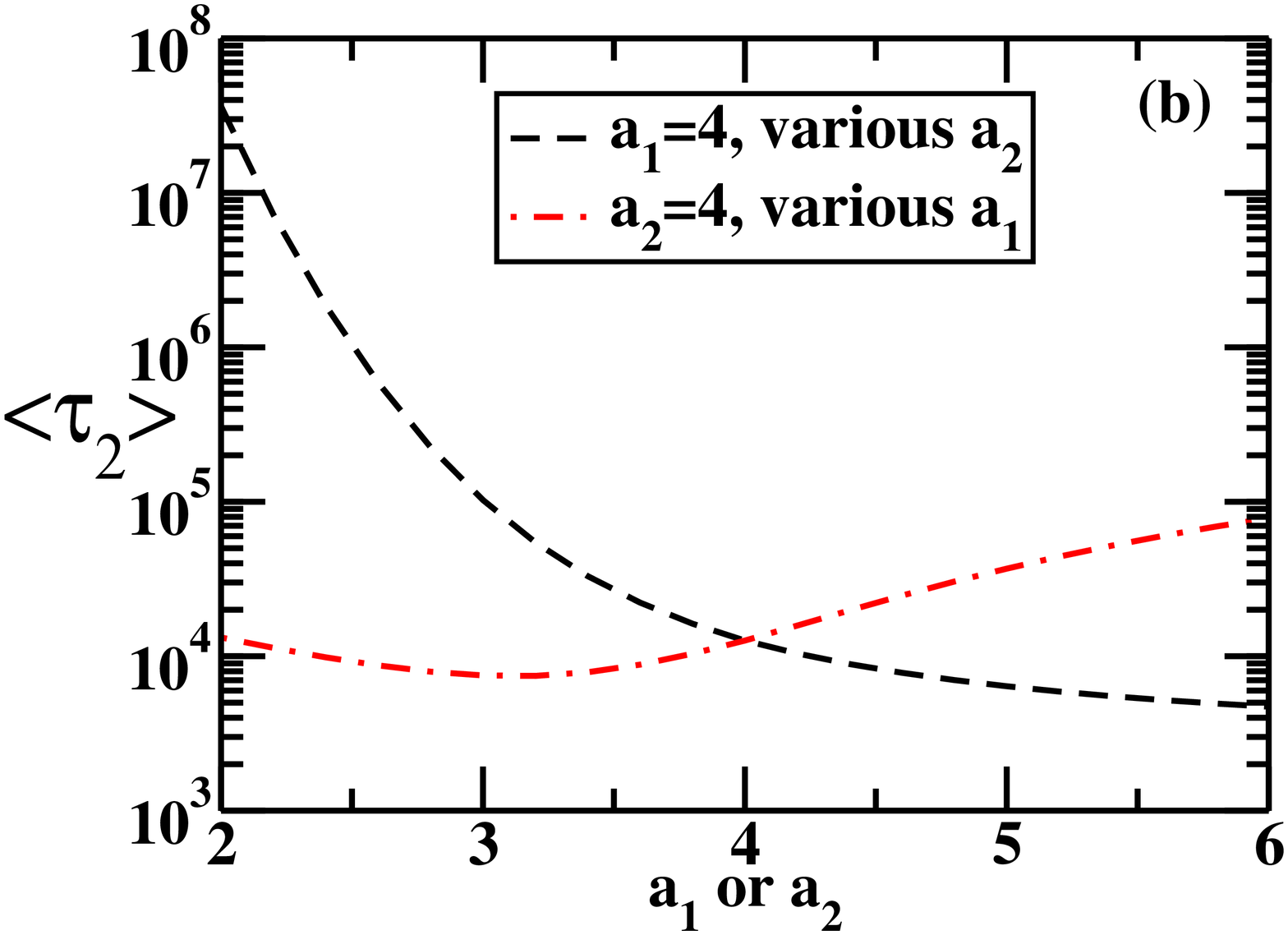} } 
\vskip 2.25em
\centering
\resizebox{70mm}{!}{\includegraphics{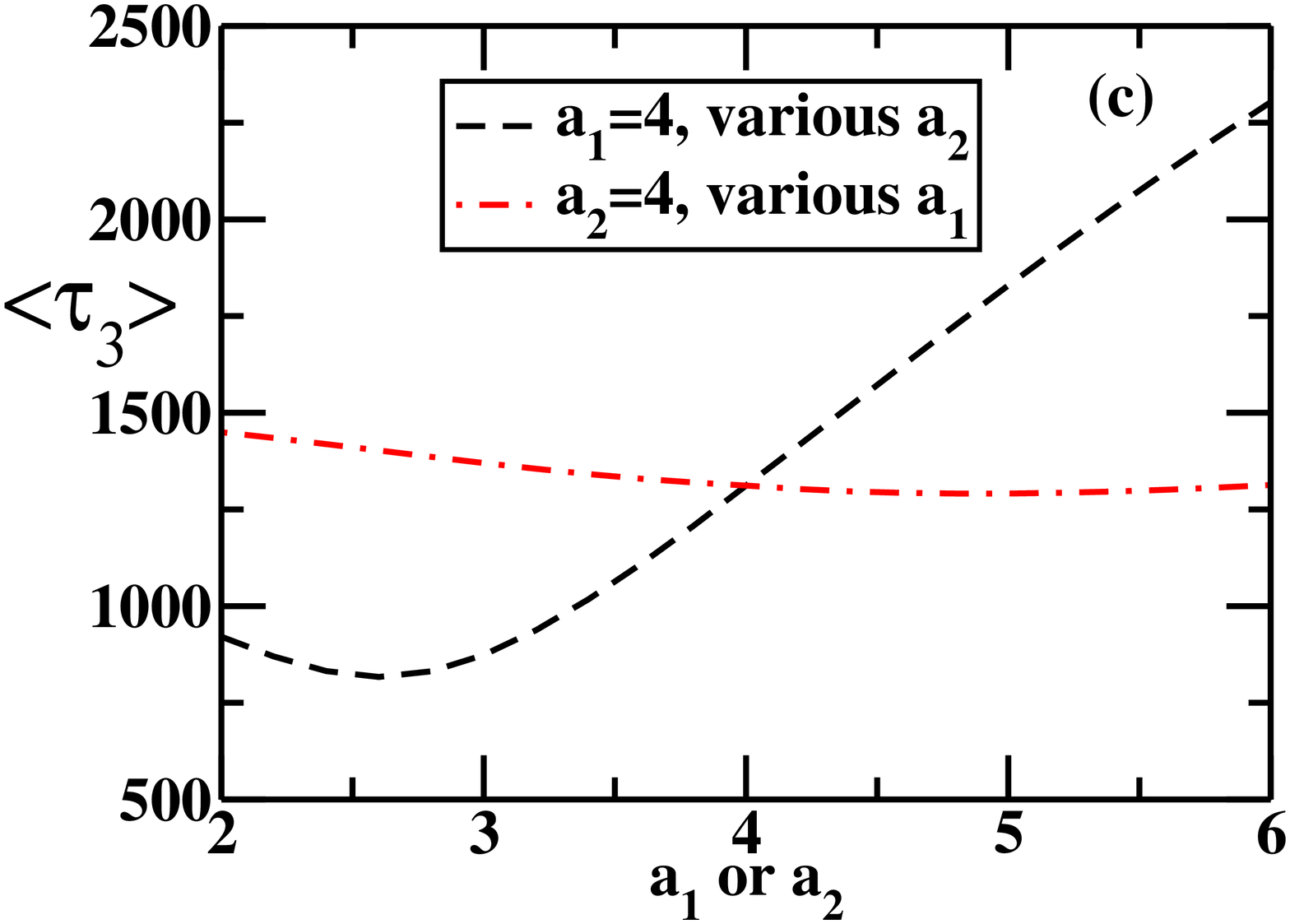} }
\caption{(Color online) Average time required for the completion of the (a) first, (b) second, and (c) third stages of translocation through an $\alpha$-hemolysin pore having an outer cylinder radius of $a_1=4$ for several values of the inner cylinder radius $a_2$ (dashed line) and vice versa (dotted-dashed line), and with $R_1=R_2=30$ and $M_1=M_2=15$, for a chain possessing $N=100$ segments.}{ \label{fig:aha4tau123} }
\end{figure}

\begin{figure}
\centering
\resizebox{83mm}{!}{\includegraphics{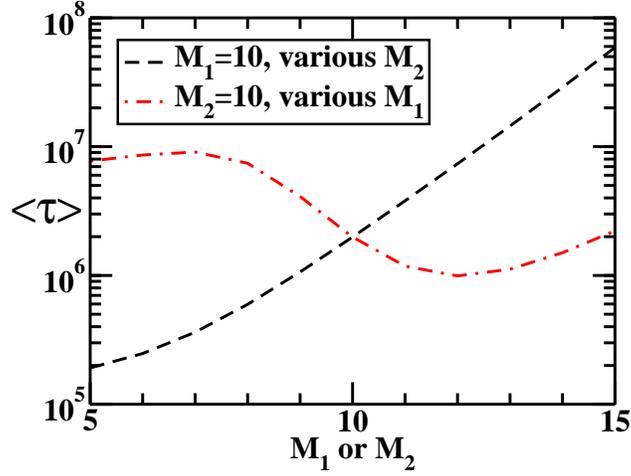} } 
\caption{(Color online) Average time of translocation from the donor to the receptor sphere through an $\alpha$-hemolysin channel with an outer cylinder length of $M_1=10$ for several values of the inner cylinder length $M_2$ (dashed line) and vice versa (dotted-dashed line), and with $R_1=R_2=30$, $a_1=4$ and $a_2=2$ for a chain possessing $N=100$ segments.}{\label{fig:ahM} }
\end{figure}

Our results are illustrated in Fig. \ref{fig:aha3a4tau}--\ref{fig:ahM} for a chain possessing $N=100$ segments. Figure \ref{fig:aha3a4tau}, which depicts the total translocation time for several values of $a_2$ with $a_1=3$ and $a_1=4$, and vice versa, and with $M_1=M_2=15$ and $R_1=R_2=30$, reveals that translocation is faster when $a_1>a_2$ than in the reverse case when the cylinder radii are interchanged so that $a_2>a_1$. This observation may be explained by the fact that chain migration into the outer cylinder from the donor sphere involves a large entropic penalty, which is reduced with increase in the outer cylinder radius.

The durations of the three stages of translocation are illustrated in Fig. \ref{fig:aha4tau123} for several values of $a_2$ with $a_1=4$, and vice versa, with $M_1=M_2=15$ and $R_1=R_2=30$. Figure \ref{fig:aha4tau123} reveals a decrease in $\left\langle \tau_1 \right\rangle$ with increase in $a_1$ for fixed $a_2$, owing to the concomitant lowering of the entropic barrier to translocation into the outer cylinder from the donor sphere. Varying $a_2$ with $a_1$ held fixed has no significant effect on $\left\langle \tau_1 \right\rangle$, and the slight dependence of $\left\langle \tau_1 \right\rangle$ on $a_2$ arises from the restriction that the leading chain end be located at the outer cylinder--inner cylinder interface within a radius of $a_2$ from the pore axis, at the end of the first stage of translocation. The second stage of translocation is hastened by an increase in $a_2$ for fixed $a_1$, owing to the reduction in the entropic cost of translocation into the inner cylinder as $a_2$ is increased. Figure \ref{fig:aha4tau123} suggests the existence of a minimum in $\left\langle \tau_2 \right\rangle$ with respect to $a_1$ for fixed $a_2$, which occurs when $a_1<a_2$. Varying $a_1$ when $a_1<a_2$ may produce two opposing effects on $\left\langle \tau_2 \right\rangle$. First, for $a_1<a_2$, the entropic gain on entering the wider inner cylinder decreases with increase in $a_1$. On the other hand, chain segments initially contained in the donor sphere must migrate across the outer cylinder before they can reach the inner cylinder, and the corresponding entropic cost is lowered as $a_1$ is increased. These opposing effects may combine to produce a minimum in $\left\langle \tau_2 \right\rangle$ with respect to $a_1$ for $a_1<a_2$, given a fixed $a_2$. When $a_1>a_2$, the loss of entropy on entering the narrower, inner cylinder grows with increase in $a_1$, leading to an increase in $\left\langle \tau_2 \right \rangle$. A slight decrease in $\left \langle \tau_3 \right \rangle$ is seen with increase in $a_1$ for fixed $a_2$. This decrease may be attributed to the fact that, as $a_1$ increases, configurations in which a greater number of chain segments is present in the outer cylinder at the beginning of the third stage, rather than in the donor sphere, become more favorable. The former configurations have a higher free energy than the latter and, hence, lead to a larger free energy drop upon complete chain transport into the receptor sphere, thus favoring translocation. A minimum in $\left \langle \tau_3 \right \rangle$ with respect to $a_2$ for fixed $a_1$ when $a_2<a_1$ is also observed. An increase in $a_2$ may influence $\left \langle \tau_3 \right \rangle$ in two opposing ways. As $a_2$ is increased, the entropy gained during the third stage of translocation into the receptor sphere is lowered. However, simultaneously, it becomes entropically more favorable for chain segments to move through the inner cylinder with increase in $a_2$. The combination of these two effects may be responsible for producing a minimum in $\left\langle \tau_3 \right\rangle$ with respect to $a_2$. The monotonic increase in $\left\langle \tau_3 \right\rangle$ with respect to increasing $a_2$ when $a_2>a_1$ may be attributed to the concomitant lowering of the entropic gain upon chain migration into the receptor sphere from the inner cylinder.

Figure \ref{fig:ahM} depicts the total translocation time as a function of $M_2$ for $M_1=10$, and vice versa, with $a_1=4$, $a_2=2$ and $R_1=R_2=30$. An increase in $M_2$ at fixed $M_1$ is accompanied by an increase in $\left\langle \tau \right \rangle$, as expected, owing to the increased cost of chain transport through the inner cylinder. The apparent maximum in $\left\langle \tau \right\rangle$ with respect to $M_1$ for fixed $M_2$ and for $M_1<M_2$ may have arisen from numerical errors in our procedure (described in the Appendix). However, an initial decrease in $\left\langle \tau \right \rangle$ with increase in $M_1$ for fixed $M_2$ may be attributed to the hastening of the second and third stages of translocation with increase in $M_1$. As $M_1$ is increased, an increasing number of chain segments is likely to be present in the outer cylinder, rather than in the donor sphere, at the end of the first stage of translocation. Such configurations provide a greater driving force for subsequent chain transport into the inner cylinder and the receptor sphere, as opposed to higher-entropy configurations with a large number of chain segments in the donor sphere. This argument is supported by the observations (not shown here) that $\left\langle \tau_3 \right\rangle$ decreases with increase in $M_1$ or $M_2$ (owing to a concomitant increase in the entropic gain upon translocation into the receptor sphere), while $\left \langle \tau_2 \right\rangle$ decreases as $M_1$ is increased and increases as $M_2$ is increased. It should also be noted that an increase in $\left\langle \tau_1 \right \rangle$ occurs as $M_1$ is increased, while $M_2$ has no effect on $\left \langle \tau_1 \right \rangle$. The increase in $\left\langle \tau_1 \right \rangle$ with respect to increasing $M_1$ may lead to an eventual increase in $\left\langle \tau \right \rangle$ as $M_1$ is increased at a fixed value of $M_2$.

\section{\label{sec:sum} Conclusions}

In this contribution, we investigate the translocation of a Gaussian chain through composite pore geometries in the absence of excluded volume effects and hydrodynamic interactions. The translocation process is modeled as the diffusion of the translocation coordinate over a free energy barrier, governed by the Fokker--Planck equation subject to the radiation boundary condition at interfaces between pore constituents. The radiation boundary condition allows the leading chain end to visit the interface many times before it finally migrates into the portion of the pore downstream of the interface.

We illustrate our method by investigating chain translocation from one spherical chamber to another through a cylindrical pore and through a two-cylinder composite pore representing an $\alpha$-hemolyin membrane channel. Our derivation of the free energy landscape is based upon the enumeration of all possible chain configurations during translocation, under the assumption that the chain length greatly exceeds the cylinder lengths. Thus, we do not consider situations wherein the chain completely enters the pore and subsequently diffuses downstream to the entrance of the receptor chamber. The latter situation may arise if the pore length greatly exceeds the polymer length, or if the pore width becomes comparable to the polymer coil size. In such situations, the time taken by the chain to diffuse across the length of the pore scales with the square of the pore length. However, we do not attempt to quantify such effects here. Moreover, our procedure for obtaining the probability densities of chain configurations within the pore disallows the chain from forming hairpin configurations at the interface between constituent parts of the pore, and is consequently restricted to small pore diameters.

Our results reveal that the time of translocation from the donor to the receptor sphere connected by a cylindrical pore monotonically increases with increase in the cylinder length. In contrast, the earlier study of Wong and Muthukumar \cite{wm}, which restricted reverse chain motions at the interface between the cylinder and the receptor sphere, predicted the occurrence of a minimum in the translocation time with respect to cylinder length. These observations suggest that reverse chain motions at the interface, which are captured by the radiation boundary condition, may contribute significantly to the total translocation time. We further establish that the translocation of a chain through an $\alpha$-hemolysin channel is faster when the chain is introduced on the \emph{cis} side of the pore, rather than the \emph{trans} side, as may be expected based on entropic considerations. This observation is consistent with the experimental results of Henrickson et al. \cite{henrickson}, who found the frequency of translocation-induced ionic-current blockades to be higher when polynucleotide chains were introduced on the \emph{cis} side, rather than the \emph{trans} side of an $\alpha$-hemolysin pore. Henrickson et al. have suggested that pore--polymer electrostatic interactions may explain the observed asymmetry. However, while pore--polymer interactions may influence the polymer concentration in the vicinity of the pore, attractive interactions serve to greatly slow the translocation process \cite{interact1, interact2}. On the other hand, the lower entropic cost of entering the pore on the \emph{cis} side and subsequent chain migration may explain the observations of Henrickson et al. Finally, the approach presented in this contribution may be applied to study translocation through other asymmetric biological or synthetic pores, including conically-shaped pores \cite{cone}, as well as other pore geometries.

\appendix
\section*{Appendix}

Translocation through a two-cylinder composite proceeds through intermediate configurations wherein the chain straddles the two-cylinder interface. Accordingly, we obtain the probability density for a chain of $N$ segments within a cylinder of radius $a$ and length $M$, whose one end is tethered at a distance of $c$ from a cylinder end surface and lies within a radial distance of $a^\prime$ from the cylinder axis, as follows:
\begin{equation} 
P_{C_{1}}(a,M,N,a^{\prime }) = \frac{8\pi caa^{\prime }}{M}\sum_{m,k=1}^{\infty }\frac{1-(-1)^{m}}{\mu
_{0,k}^{2}}\frac{J_{1}\left( \mu _{0,k}\frac{a^{\prime }}{a}\right) }{%
J_{1}\left( \mu _{0,k}\right) }\exp \left[ -\left( \frac{\mu _{0,k}^{2}}{%
a^{2}}+\frac{m^{2}\pi ^{2}}{M^{2}}\right) \frac{N}{6} \right] \label{eq:ahPC1}
\end{equation}
Similarly, the probability density for a chain within the cylinder with both ends tethered at distances of $c_1$ and $c_2$ from the cylinder end surfaces, and lying within radial distances of $a^\prime$ and $a^{\prime\prime}$ from the cylinder axis, is
\begin{multline}
P_{C_{2}}(a,M,N,a^{\prime },a^{\prime \prime })=\frac{8\pi ^{3}a^{\prime
}a^{\prime \prime }c_{1}c_{2}}{M^{3}}\sum_{m,k=1}^{\infty }\frac{m^{2}}{\mu
_{0,k}^{2}}\left( -1\right) ^{m+1}\frac{J_{1}\left( \mu _{0,k}\frac{%
a^{\prime }}{a}\right) J_{1}\left( \mu _{0,k}\frac{a^{\prime \prime }}{a}%
\right) }{J_{1}^{2}\left( \mu _{0,k}\right) } \\
\times \exp \left[ -\left( \frac{\mu _{0,k}^{2}}{a^{2}}+\frac{m^{2}\pi ^{2}}{M^{2}}%
\right) \frac{N}{6} \right] \label{eq:ahPC2}
\end{multline}
We approximate the probability density for a chain whose one end lies within the outer cylinder and the other end within the inner cylinder with the product $P_{C_1} \left( a_1,M_1,N_1,a_{\min }\right) P_{C_1} \left(a_{2},M_{2},N_{2},a_{\min }\right) / (\pi a_{\min }^{2})$, where 
$a_{\min} = \min{(a_1, a_2)}$. Because Eqs. (\ref{eq:ahPC1}) and (\ref{eq:ahPC2}) involve averaging over all possible positions of the chain segment at the interface between the cylinders, our results do not reduce exactly to the results of Sec. \ref{sec:cylinder} upon setting $a_1=a_2$. However, we have verified that our approximations qualitatively reproduce the behavior of the translocation time obtained from Sec. \ref{sec:cylinder} for the case $a_1=a_2$.

With the aid of the above definitions, the free energies of the chain configurations illustrated in Fig. \ref{fig:ahallpics} are defined by the following expressions:
\begin{equation}
F_{(a)}(l=0)=-\ln \left[ P_{S_{1}}\left( R_{1},N\right) \pi a_{1}^{2}\right]
\end{equation}
\begin{equation}
F_{(b)}(l<R_l)=-\ln \left[ P_{S_{1}}\left( R_{1},N-l\right) P_{C_{1}}\left(
a_{1},M_{1},l,a_{1}\right) \right]
\end{equation}
\begin{equation}
F_{(c)}(l=R_{l}<N)=-\ln \left[ P_{S_{1}}\left( R_{1},N-l\right)
P_{C_{2}}\left( a_{1},M_{1},l,a_{1},a_{\min }\right) \right]
\end{equation}
\begin{equation}
F_{(d)}(l=R_l=N)=-\ln \left[ P_{C_{2}}\left( a_{1},M_{1},N,a_{1},a_{\min
}\right) \right]
\end{equation}
\begin{multline}
 F_{(e)}(m<R_m, R_l<N-1)=-\ln \left[ P_{S_{1}}\left( R_{1},N-R_{l}-m\right)
P_{C_{2}}\left( a_{1},M_{1},R_{l},a_{1},a_{\min }\right) \right. \\
\times \left. P_{C_{1}}\left(
a_{2},M_{2},m,a_{\min }\right) /\left( \pi a_{\min }^{2}\right) \right]
\end{multline}
\begin{multline}
F_{(f)}(m=R_{m},R_{m}<N-R_{l},R_{l}< N-1)= -\ln \left[ P_{S_{1}}\left( R_{1},N-R_{l}-R_{m}\right) \right. \\
\times \left. P_{C_{2}}\left( a_{1},M_{1},R_{l},a_{1},a_{\min }\right) P_{C_{2}}\left(
a_{2},M_{2},R_{m},a_{\min },a_{2}\right) /\left( \pi a_{\min }^{2}\right) %
\right]
\end{multline}
\begin{multline}
F_{(g)}(m=R_m=N-R_l, R_l<N)=-\ln \left[ P_{C_{2}}\left( a_{1},M_{1},R_{l},a_{1},a_{\min
}\right) \right. \\
\times \left. P_{C_{2}}\left( a_{2},M_{2},R_{m},a_{\min },a_{2}\right) /\left(
\pi a_{\min }^{2}\right) \right]
\end{multline}
\begin{equation}
 F_{(h)}(m=N-R_l)=-\ln \left[ P_{C_{2}}\left( a_{1},M_{1},R_{l},a_{1},a_{\min
}\right) P_{C_{1}}\left( a_{2},M_{2},N-R_{l},a_{\min }\right) /\left( \pi
a_{\min }^{2}\right) \right]
\end{equation}
\begin{equation}
 F_{(i)}(N-R_{l}<m<R_{m})=-\ln \left[ P_{C_{1}}\left( a_{1},M_{1},N-m,a_{\min }\right)
P_{C_{1}}\left( a_{2},M_{2},m,a_{\min }\right) /\left( \pi a_{\min
}^{2}\right) \right]
\end{equation}
\begin{equation}
 F_{(j)}(m=R_m<N)=-\ln \left[ P_{C_{1}}\left( a_{1},M_{1},N-m,a_{\min }\right)
P_{C_{2}}\left( a_{2},M_{2},m,a_{\min },a_{2}\right) /\left( \pi a_{\min
}^{2}\right) \right]
\end{equation}
\begin{equation}
 F_{(k)}(m=R_m=N)=-\ln \left[ P_{C_{2}}\left( a_{2},M_{2},N,a_{\min },a_{2}\right) %
\right]
\end{equation}
\begin{multline}
 F_{(l)}(1\leq n<N-R_l-R_m)=-\ln \left[
 P_{S_{1}}\left( R_{1},N-R_{l}-R_{m}-n\right) P_{C_{2}}\left(
a_{1},M_{1},R_{l},a_{1},a_{\min }\right)   \right. \\
\times \left. P_{C_{2}}\left(a_{2},M_{2},R_{m},a_{\min },a_{2}\right) P_{S_{1}} \left( R_{2},n\right)
/  \left( \pi a_{\min }^{2}\right) \right]
\end{multline}
\begin{multline}
 F_{(m)}(n=N-R_{l}-R_{m})=-\ln \left[ P_{C_{2}}\left( a_{1},M_{1},R_{l},a_{1},a_{\min
}\right) P_{C_{2}}\left( a_{2},M_{2},R_{m},a_{\min },a_{2}\right) \right. \\
\times \left. P_{S_{1}}\left( R_{2},n\right) /\left( \pi a_{\min }^{2}\right) \right]
\end{multline}
\begin{multline}
 F_{(n)}(N-R_{l}-R_{m}<n<N-R_{m})= -\ln \left[ P_{C_{1}}\left( a_{1},M_{1},N-R_{m}-n,a_{\min
}\right) \right. \\
\times \left. P_{C_{2}}\left( a_{2},M_{2},R_{m},a_{\min },a_{2}\right)
P_{S_{1}}\left( R_{2},n\right) /\left( \pi a_{\min }^{2}\right) \right]
\end{multline}
\begin{equation}
 F_{(o)}(n=N-R_m)=-\ln \left[ P_{C_{2}}\left( a_{2},M_{2},R_{m},a_{\min
},a_{2}\right) P_{S_{1}}\left( R_{2},n\right) \right]
\end{equation}
\begin{equation}
 F_{(p)}(N-R_{m}<n<N)=-\ln \left[ P_{C_{1}}\left( a_{2},M_{2},N-n,a_{2}\right)
P_{S_{1}}\left( R_{2},n\right) \right]
\end{equation}
\begin{equation}
 F_{(q)}(n=N)=-\ln \left[ P_{S_{1}}\left( R_{2},N\right) \pi a_{2}^{2}\right]
\end{equation}

We identify the probabilities $P(R_l)$ and $P(R_m|R_l)$ employed in Eqs. (\ref{eq:tau1ah})--(\ref{eq:tau3ah}) with the weights of the corresponding chain configurations, as given by the expressions:
\begin{equation}
P\left( R_{l}\right) = 
\begin{cases}
P_{S_{1}}\left( R_{1},N-R_{l}\right) P_{C_{2}}\left(
a_{1},M_{1},R_{l},a_{1},a_{\min }\right), \ M_{1} \leq R_{l} \leq N-1 \\
P_{C_{2}}\left( a_{1},M_{1},N,a_{1},a_{\min }\right), \ R_{l}=N
\end{cases}
\end{equation}%
and 
\begin{equation}
P(R_{m}|R_{l}) = 
\begin{cases}
P_{S_{1}}\left( R_{1},N-R_{l}-R_{m}\right) P_{C_{2}}\left(
a_{1},M_{1},R_{l},a_{1},a_{\min }\right) \\
\qquad \qquad \times P_{C_{2}}\left(
a_{2},M_{2},R_{m},a_{\min },a_{2}\right) /\left( \pi a_{\min }^{2}\right), \
M_{1} \leq R_{l}\leq N-1,R_{m}<N-R_{l} \\
P_{C_{2}}\left( a_{1},M_{1},R_{l},a_{1},a_{\min }\right)
P_{C_{2}}\left(a_{2},M_{2},R_{m},a_{\min },a_{2}\right) /\left( \pi a_{\min }^{2}\right), \\ 
\qquad \qquad M_{1} \leq  R_{l}\leq N-1,R_{m}=N-R_{l}  \\
P_{C_{1}}\left( a_{1},M_{1},N-R_{m},a_{\min }\right) P_{C_{2}}\left(
a_{2},M_{2},R_{m},a_{\min },a_{2}\right) /\left( \pi a_{\min }^{2}\right), \\
\qquad \qquad M_{1} \leq  R_{l}\leq N-1,N-R_{l}<R_{m}<N \text{ or } R_{l}=N,R_{m}<N  \\
P_{C_{2}}\left( a_{2},M_{2},R_{m},a_{\min },a_{2}\right), \ M_{1}\leq R_{l}\leq N,R_{m}=N
\end{cases}
\end{equation}

The rate constants $k_1$ and $k_2$ appearing in the radiation boundary conditions in Eqs. (\ref{eq:k1}) and (\ref{eq:k2}) may be obtained by means of the following expressions:
\begin{equation}
\frac{\partial F\left( l\right) }{\partial l}=
\begin{cases}
-\ln \left[ P_{s_{1}}\left(
R_{1},N-R_{l}\right) P_{C_{2}}\left( a_{1},M_{1},R_{l},a_{1},a_{\min
}\right) \right]  \\
\qquad \qquad + \ln \left[ P_{s_{1}}\left( R_{1},N-R_{l}+1\right) P_{C_{1}}\left(
a_{1},M_{1},R_{l}-1,a_{1}\right) \right], \ R_{l}<N \\
-\ln \left[ P_{C_{2}}\left( a_{1},M_{1},N,a_{1},a_{\min }\right) \right]
+ \ln \left[ P_{s_{1}}\left( R_{1},1\right) P_{C_{1}}\left(
a_{1},M_{1},N-1,a_{1}\right) \right], \\
\qquad \qquad R_{l}=N
\end{cases}
\end{equation}

\begin{equation}
\frac{\partial F\left( m\right) }{\partial m} =
\begin{cases}
-\ln \left[ P_{S_{1}}\left(
R_{1},N-R_{l}-R_{m}\right) P_{C_{2}}\left( a_{1},M_{1},R_{l},a_{1},a_{\min
}\right) P_{C_{2}}\left( a_{2},M_{2},R_{m},a_{\min },a_{2}\right) /\left(
\pi a_{\min }^{2}\right) \right] \\
\qquad \qquad + \ln \left[ P_{S_{1}}\left( R_{1},N-R_{l}-R_{m}+1\right) P_{C_{2}}\left(
a_{1},M_{1},R_{l},a_{1},a_{\min }\right) \right.  \\
\qquad \qquad  \times \left. P_{C_{1}}\left(
a_{2},M_{2},R_{m}-1,a_{\min }\right) /\left( \pi a_{\min }^{2}\right) \right],
R_{m}<N-R_{l} \\
-\ln \left[ P_{C_{2}}\left( a_{1},M_{1},R_{l},a_{1},a_{\min }\right) 
P_{C_{2}}\left( a_{2},M_{2},R_{m},a_{\min },a_{2}\right) /\left( \pi a_{\min
}^{2}\right) \right]  \\
\qquad \qquad + \ln \left[ P_{S_{1}}\left( R_{1},1\right) P_{C_{2}}\left(
a_{1},M_{1},R_{l},a_{1},a_{\min }\right) P_{C_{1}}\left(
a_{2},M_{2},R_{m}-1,a_{\min }\right) /\left( \pi a_{\min }^{2}\right) \right], \\
\qquad \qquad  R_{m}=N-R_{l} \\
-\ln \left[ P_{C_{1}}\left( a_{1},M_{1},N-R_{m},a_{\min }\right)
P_{C_{2}}\left( a_{2},M_{2},R_{m},a_{\min },a_{2}\right) /\left( \pi a_{\min
}^{2}\right) \right]  \\
\qquad \qquad + \ln \left[ P_{C_{1}}\left( a_{1},M_{1},N-R_{m}+1,a_{\min }\right)
P_{C_{1}}\left( a_{2},M_{2},R_{m}-1,a_{\min }\right) /\left( \pi a_{\min
}^{2}\right) \right], \\
\qquad \qquad N-R_{l}<R_{m}<N \\
-\ln \left[ P_{C_{2}}\left( a_{2},M_{2},N,a_{\min },a_{2}\right) \right]
+\ln \left[ P_{C_{1}}\left( a_{1},M_{1},1,a_{\min }\right)  \right. \\
\qquad \qquad \times \left. P_{C_{1}}\left(
a_{2},M_{2},N-1,a_{\min }\right) /\left( \pi a_{\min }^{2}\right) \right],\  
R_{m}=N
\end{cases}
\end{equation}

\begin{equation}
W_l(R_{l}) \sim
\begin{cases}
P_{S_{1}}\left( R_{1},N-R_{l}\right) P_{C_{2}}\left(
a_{1},M_{1},R_{l},a_{1},a_{\min }\right), \ R_{l}<N \\
P_{C_{2}}\left( a_{1},M_{1},N,a_{1},a_{\min }\right),\  R_{l}=N
\end{cases}
\end{equation}

\begin{equation}
W_m(R_{l},R_{m})\sim
\begin{cases} 
P_{S_{1}}\left( R_{1},N-R_{l}-R_{m}\right)
P_{C_{2}}\left( a_{1},M_{1},R_{l},a_{1},a_{\min }\right) \\
\qquad \qquad \times P_{C_{2}}\left(a_{2},M_{2},R_{m},a_{\min },a_{2}\right) /\left( \pi a_{\min }^{2}\right), \\
\qquad \qquad  R_{m} < N-R_{l} \\
P_{C_{2}}\left( a_{1},M_{1},R_{l},a_{1},a_{\min }\right) P_{C_{2}}\left(
a_{2},M_{2},R_{m},a_{\min },a_{2}\right) /\left( \pi a_{\min }^{2}\right),  \\
\qquad \qquad R_{m} = N-R_{l}  \\
P_{C_{1}}\left( a_{1},M_{1},N-R_{m},a_{\min }\right) P_{C_{2}}\left(
a_{2},M_{2},R_{m},a_{\min },a_{2}\right) /\left( \pi a_{\min }^{2}\right), \\
\qquad \qquad  N-R_{l}<R_{m}<N  \\
P_{C_{2}}\left( a_{2},M_{2},N,a_{\min },a_{2}\right),\ R_{m}=N
\end{cases}
\end{equation}

\begin{equation}
\frac{\partial W_l }{\partial l}\sim
\begin{cases}
 P_{s_{1}}\left(
R_{1},N-R_{l}\right) P_{C_{2}}\left( a_{1},M_{1},R_{l},a_{1},a_{\min
}\right) \\
\qquad \qquad - P_{s_{1}}\left( R_{1},N-R_{l}+1\right) P_{C_{1}}\left(
a_{1},M_{1},R_{l}-1,a_{1}\right), \ R_{l}<N \\
 P_{C_{2}}\left( a_{1},M_{1},N,a_{1},a_{\min }\right)-
 P_{S_{1}}\left( R_{1},1\right) P_{C_{1}}\left(
a_{1},M_{1},N-1,a_{1}\right),\  R_{l}=N
\end{cases}
\end{equation}
and
\begin{equation}
\frac{\partial W_m }{\partial m} \sim
\begin{cases}
P_{S_{1}}\left(R_{1},N-R_{l}-R_{m}\right) P_{C_{2}}\left( a_{1},M_{1},R_{l},a_{1},a_{\min
}\right)   P_{C_{2}}\left( a_{2},M_{2},R_{m},a_{\min },a_{2}\right) /\left(
\pi a_{\min }^{2}\right)  \\
\qquad \qquad - P_{S_{1}}\left( R_{1},N-R_{l}-R_{m}+1\right) P_{C_{2}}\left(
a_{1},M_{1},R_{l},a_{1},a_{\min }\right)  P_{C_{1}}\left(
a_{2},M_{2},R_{m}-1,a_{\min }\right) \\
\qquad \qquad  /\left( \pi a_{\min }^{2}\right), R_{m}<N-R_{l} \\
 P_{C_{2}}\left( a_{1},M_{1},R_{l},a_{1},a_{\min }\right) 
P_{C_{2}}\left( a_{2},M_{2},R_{m},a_{\min },a_{2}\right) /\left( \pi a_{\min
}^{2}\right)  \\
\qquad \qquad - P_{S_{1}}\left( R_{1},1\right) P_{C_{2}}\left(
a_{1},M_{1},R_{l},a_{1},a_{\min }\right) P_{C_{1}}\left(
a_{2},M_{2},R_{m}-1,a_{\min }\right) /\left( \pi a_{\min }^{2}\right), \\
\qquad \qquad  R_{m}=N-R_{l} \\
P_{C_{1}}\left( a_{1},M_{1},N-R_{m},a_{\min }\right)
P_{C_{2}}\left( a_{2},M_{2},R_{m},a_{\min },a_{2}\right) /\left( \pi a_{\min
}^{2}\right) \\
\qquad \qquad - P_{C_{1}}\left( a_{1},M_{1},N-R_{m}+1,a_{\min }\right)
P_{C_{1}}\left( a_{2},M_{2},R_{m}-1,a_{\min }\right) /\left( \pi a_{\min
}^{2}\right), \\
\qquad \qquad N-R_{l}<R_{m}<N \\
 P_{C_{2}}\left( a_{2},M_{2},N,a_{\min },a_{2}\right) -
P_{C_{1}}\left( a_{1},M_{1},1,a_{\min }\right) 
P_{C_{1}}\left(
a_{2},M_{2},N-1,a_{\min }\right) /\left( \pi a_{\min }^{2}\right),\\
\qquad \qquad R_{m}=N
\end{cases}
\end{equation}

\section*{Acknowledgments}

The authors would like to acknowledge support from the Welch Foundation (Grants No. C-1559 and C-1668) and the U.S. National Science Foundation (Grants No. CHE-0237105, ECCS-0708765 and EEC-0647452).

\bibliography{pore_paper}

\clearpage
\pagebreak

\clearpage
\pagebreak

\clearpage
\pagebreak

\clearpage
\pagebreak

\clearpage
\pagebreak

\clearpage
\pagebreak

\clearpage
\pagebreak

\clearpage
\pagebreak

\clearpage
\pagebreak
\listoffigures

\end{document}